\pdfoutput=1

\documentclass[11pt]{article}

\usepackage{acl2023}
\usepackage[utf8]{inputenc} 
\usepackage[T1]{fontenc}    
\usepackage{hyperref}       
\usepackage{url}            
\usepackage{booktabs,tabularx}      
\usepackage{amsfonts}       
\usepackage{nicefrac}       
\usepackage{microtype}      
\usepackage{xcolor}         
\usepackage{float}          
\usepackage{booktabs, graphicx, multirow}
\usepackage{amsmath,amssymb}
\usepackage{array,colortbl}
\usepackage{tikz}             
\usetikzlibrary{plotmarks}  
\usepackage{subcaption}     
\usepackage{amsthm}
\usetikzlibrary{shapes.geometric, shapes.symbols, arrows.meta, positioning, calc, backgrounds, shadows, fit, decorations, decorations.pathmorphing}
\usepackage{fancyvrb}
\usepackage{fvextra}
\usepackage[most]{tcolorbox}
\tcbuselibrary{skins,breakable}
\usepackage{algorithm} 
\usepackage{algorithmic}
\usepackage{seqsplit}          
\usepackage[normalem]{ulem}
\usepackage{enumitem}
\usepackage{soul}
\usepackage{booktabs,multirow,pifont}
\usepackage[table]{xcolor}

\definecolor{promptframe}{HTML}{3D4A5C}
\definecolor{promptback}{HTML}{F6F7F9}
\definecolor{slotink}{HTML}{0B6E4F}

\colorlet{pubcol}{green!55!black}   
\colorlet{privcol}{red!65}          
\colorlet{noisecol}{orange!85!black}
\colorlet{graycol}{gray!70}
\definecolor{cigreen}{RGB}{22,163,74}

\newtcolorbox[auto counter, number within=section]{promptbox}[2][]{%
  enhanced, breakable,
  colback=promptback, colframe=promptframe,
  boxrule=0.4pt, arc=1.5pt,
  boxsep=0pt, left=2.5pt, right=2.5pt, top=2.5pt, bottom=2.5pt,
  fonttitle=\bfseries\scriptsize, coltitle=white,
  title={Prompt~\thetcbcounter: #2},
  #1
}

\DefineVerbatimEnvironment{promptverb}{Verbatim}{%
  commandchars=!\[\],
  fontsize=\fontsize{6.2}{7}\selectfont,
  breaklines=true, breakanywhere=true,
  breaksymbolleft={}, breaksymbolright={},
  breakautoindent=false, breakindent=0pt,
  xleftmargin=0pt, numbers=none
}
 
\fvset{fontsize=\small, commandchars=\!\[\], xleftmargin=0pt}

\definecolor{brickred}{rgb}{0.8, 0.25, 0.33}
\definecolor{burntorange}{rgb}{0.8, 0.33, 0.0}

\newcommand{\cmark}{\textcolor{green!50!black}{\ding{51}}}
\newcommand{\xmark}{\textcolor{red!70!black}{\ding{55}}}
\newcommand{\pmark}{\textcolor{orange}{$\sim$}}

\newtheorem{theorem}{Theorem}

\newtheorem{definition}{Definition}

\newcommand{\A}{\mathcal{A}}
\newcommand{\X}{\mathcal{X}}

\newcommand{\Apub}{\A_{\mathrm{pub}}}
\newcommand{\Apriv}{\A_{\mathrm{priv}}}

\newcommand{\Ext}{\textsc{Ext}}
\renewcommand{\epsilon}{\varepsilon}   
\newcommand{\Classify}{\textsc{Classify}}
\newcommand{\Priv}{\textsc{Priv}}
\newcommand{\sdgci}{\textsc{SDG-CI}}
\newcommand{\anci}{\textsc{ANON-CI}}

\title{Who's Asking? \\ Audience-Conditioned Privatization}

\author{%
  Daniil Filienko\thanks{\ Work done while interning at Zillow.} \\
  University of Washington Tacoma\\
  daniilf@uw.edu \\
  \And
  Anusha Bagalkotkar \\
  Zillow \\
  anushaba@zillowgroup.com \\
  \AND
  Xan Vongsathorn \\
  Zillow \\
  xanv@zillowgroup.com \\
  \And
  Sikha Pentyala \\
  University of Washington Tacoma \\
  sikha@uw.edu \\
}

\begin{document}

\maketitle

\begin{abstract}
    The abundance of user-generated conversational data makes the question of how to responsibly derive insights from confidential data without putting the individuals in it at risk ever more important. However, the majority of existing differentially private (DP) synthetic text generation (SDG) and anonymization approaches ignore the context of the data's downstream usage, which results in loss of important information. DP-SDG provides the same guarantee regardless of the recipient, and anonymization relies on predefined attribute categories. They both ignore the observation that an attribute that is harmless to one audience can be sensitive to another, and each audience typically needs only some attributes preserved. 
    We propose Ext-Classify-Priv, a framework that (1) extracts the information types present in a corpus, (2) classifies which of them are private for a given audience following contextual integrity, and (3) protects only those attributes, using either anonymization or SDG with Feature-DP approaches. On Reddit data with 3 audiences, our audience-specific releases are more faithful than a single audience-agnostic release for every audience. On bioRxiv data, our SDG method improves MAUVE from 0.097 to 0.420 at $\varepsilon=1$ and on the proprietary real estate data, our anonymization method improves MAUVE from 0.698 to 0.959 over the best baselines.
\end{abstract}

\section{Introduction}
Organizations hold large volumes of user-generated text, particularly in sensitive domains such as healthcare and real estate. This data can be used to derive valuable insights. For example, within an organization, product teams can learn where their systems fall short, and market research teams can track emerging customer needs. However, such text routinely contains personal and professional details, so it cannot be shared broadly with the stakeholders who would benefit from it. Even releasing insights derived from the text, rather than the text itself, can let an adversary infer the contents of individual records \cite{annamalai2026cliopatra}. As a result, much of this data goes unused. Putting it to use without exposing the individuals who contributed it is a key open problem, both for analyzing deployed systems responsibly and for building new solutions \cite{liu2025urania}.

In practice, the same data is requested by many stakeholders, and there is no single answer to what must be protected. Audiences differ in what they need from the data and in what they are allowed to learn about the people in it. For example, consider conversations between a company's chatbot and its users. Fairness evaluators checking that the chatbot treats different user groups equally need user's characteristics. The product team measuring user satisfaction needs the topic and outcome of each conversation, but not who the user is. Data holders therefore face a choice between applying the strictest protection to every release, which destroys the insights each audience needs, and not using the data at all.

Existing privatization methods do not resolve this choice, because they produce the same release for every audience. Differentially private (DP) \cite{dwork2006calibrating} synthetic data generation (SDG) \cite{xiedifferentially, chien2026maple} protects every attribute of every record equally, regardless of who receives the data \cite{chien2026maple, liu2025urania}. LLM-based anonymizers rewrite records to hide a predefined list of attributes that is fixed for all audiences \cite{rupta, staab2025language}.

%
%

\providecommand{\drecq}[8]{%
  \node[#8] (#1) at (#2,#3) {};
  \draw[#4] ([xshift=-0.38cm,yshift=0.18cm]#1.center) -- ([xshift=0.38cm,yshift=0.18cm]#1.center);
  \draw[#5] ([xshift=-0.38cm,yshift=0.06cm]#1.center) -- ([xshift=0.38cm,yshift=0.06cm]#1.center);
  \draw[#6] ([xshift=-0.38cm,yshift=-0.06cm]#1.center) -- ([xshift=0.38cm,yshift=-0.06cm]#1.center);
  \draw[#7] ([xshift=-0.38cm,yshift=-0.18cm]#1.center) -- ([xshift=0.38cm,yshift=-0.18cm]#1.center);
}
\providecommand{\scol}[3]{%
  \node[#3head] at ([xshift={(#2-2.5)*0.40cm},yshift=0.39cm]#1.center) {$a_{#2}$};
  \node[#3cell] at ([xshift={(#2-2.5)*0.40cm},yshift=0.13cm]#1.center) {};
  \node[#3cell] at ([xshift={(#2-2.5)*0.40cm},yshift=-0.13cm]#1.center) {};
  \node[#3cell] at ([xshift={(#2-2.5)*0.40cm},yshift=-0.39cm]#1.center) {};
}
\providecommand{\stab}[7]{%
  \node[minimum width=1.60cm,minimum height=1.04cm,inner sep=0pt] (#1) at (#2,#3) {};
  \scol{#1}{1}{#4}\scol{#1}{2}{#5}\scol{#1}{3}{#6}\scol{#1}{4}{#7}%
}

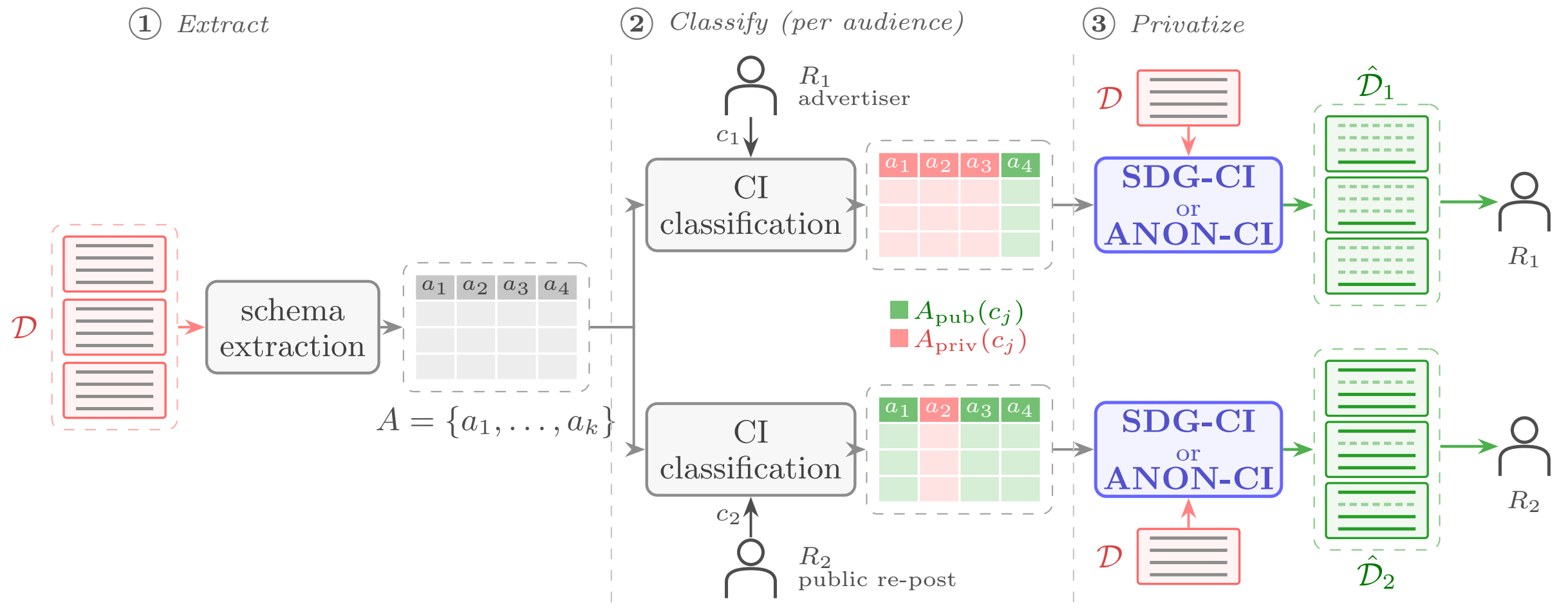
\begin{figure*}[t]
\centering
\colorlet{privcol}{red!70}
\colorlet{mechcol}{blue!60}
\colorlet{pubcol}{green!55!black}
\colorlet{audcol}{black!70}
\resizebox{\textwidth}{!}{%
\begin{tikzpicture}[
  >={Stealth[length=2mm]},
  font=\small,
  audience/.pic={
    \node[inner sep=0pt,minimum width=0.50cm,minimum height=0.60cm] (-box) at (0,0.03) {};
    \draw[audcol,fill=white,line width=0.9pt]
      (-0.24,-0.26) -- (-0.24,-0.10)
      .. controls (-0.24,0.00) and (-0.14,0.03) .. (0,0.03)
      .. controls (0.14,0.03) and (0.24,0.00) .. (0.24,-0.10)
      -- (0.24,-0.26);
    \draw[audcol,fill=white,line width=0.9pt] (0,0.19) circle (0.12);
  },
  privrec/.style={draw=privcol!85,fill=privcol!7,rounded corners=0.8pt,
                  minimum width=1.0cm,minimum height=0.54cm,inner sep=0pt,line width=0.6pt},
  relrec/.style={draw=pubcol!85,fill=pubcol!6,rounded corners=0.8pt,
                  minimum width=1.0cm,minimum height=0.54cm,inner sep=0pt,line width=0.6pt},
  neu/.style={draw=black!45,line width=0.9pt},
  keep/.style={draw=pubcol!85,line width=0.9pt},
  prot/.style={draw=pubcol!40,line width=0.9pt,dash pattern=on 2pt off 1.2pt},
  cell/.style={minimum width=0.40cm,minimum height=0.26cm,inner sep=0pt,
               draw=white,line width=0.6pt,font=\tiny},
  neuhead/.style={cell,fill=black!22,text=black!75},
  neucell/.style={cell,fill=black!7},
  pubhead/.style={cell,fill=pubcol!55,text=white},
  pubcell/.style={cell,fill=pubcol!16},
  senhead/.style={cell,fill=privcol!60,text=white},
  sencell/.style={cell,fill=privcol!16},
  grp/.style={draw=black!35,dashed,rounded corners=3pt,inner sep=3pt},
  stepbox/.style={draw=black!45,fill=black!3,rounded corners=3pt,line width=0.8pt,
               text=black!75,align=center,font=\footnotesize,
               minimum width=1.5cm,minimum height=0.9cm},
  mech/.style={draw=mechcol,fill=mechcol!8,rounded corners=3pt,line width=0.9pt,
               text=mechcol!80!black,align=center,font=\footnotesize,
               minimum width=1.75cm,minimum height=0.9cm,inner sep=2pt},
  inar/.style={->,draw=privcol!70,line width=0.7pt},
  outar/.style={->,draw=pubcol!70,line width=0.8pt},
  neuar/.style={->,draw=black!45,line width=0.7pt},
  audar/.style={->,draw=audcol,line width=0.7pt},
  lbl/.style={font=\scriptsize\itshape},
  audlbl/.style={font=\scriptsize,text=audcol,align=left},
  dlbl/.style={text=privcol!85!black,font=\small},
  stepnum/.style={circle,draw=black!55,inner sep=0.8pt,font=\scriptsize\bfseries,
                  text=black!70}
]

\drecq{r1}{0}{0.62}{neu}{neu}{neu}{neu}{privrec}
\drecq{r2}{0}{0}{neu}{neu}{neu}{neu}{privrec}
\drecq{r3}{0}{-0.62}{neu}{neu}{neu}{neu}{privrec}
\node[grp,draw=privcol!45,fit=(r1)(r2)(r3),
      label={[dlbl]left:$\mathcal{D}$}] (D) {};

\node[stepbox] (E) at (1.75,0) {schema\\extraction};
\draw[inar] (D.east) -- (E.west);

\stab{S}{3.75}{0}{neu}{neu}{neu}{neu}
\node[grp,fit=(S)] (Sg) {};
\node[font=\footnotesize,below=0pt of Sg,text=black!70] {$A=\{a_1,\dots,a_k\}$};
\draw[neuar] (E.east) -- (Sg.west);

\node[stepbox] (CIa) at (6.25,1.2)  {CI\\classification};
\node[stepbox] (CIb) at (6.25,-1.2) {CI\\classification};
\draw[neuar] (Sg.east) -- (5.10,0) |- (CIa.west);
\draw[neuar] (Sg.east) -- (5.10,0) |- (CIb.west);

\pic (ua) at (6.25,2.34)   {audience};
\pic (ub) at (6.25,-2.40)  {audience};
\node[audlbl,anchor=west] at ([xshift=2pt]ua-box.east) {$R_1$\\[-2pt]{\tiny advertiser}};
\node[audlbl,anchor=west] at ([xshift=2pt]ub-box.east) {$R_2$\\[-2pt]{\tiny public re-post}};
\draw[audar] (ua-box.south) -- node[font=\scriptsize,left,text=audcol,inner sep=2pt] {$c_1$} (CIa.north);
\draw[audar] (ub-box.north) -- node[font=\scriptsize,left,text=audcol,inner sep=2pt] {$c_2$} (CIb.south);

\stab{T1}{8.30}{1.2}{sen}{sen}{sen}{pub}
\node[grp,fit=(T1)] (T1g) {};
\stab{T2}{8.30}{-1.2}{pub}{sen}{pub}{pub}
\node[grp,fit=(T2)] (T2g) {};
\draw[neuar] (CIa.east) -- (T1g.west);
\draw[neuar] (CIb.east) -- (T2g.west);
\node[font=\scriptsize,align=left,inner sep=0pt] at (8.30,0)
  {\textcolor{pubcol!55}{\rule{5pt}{5pt}}\,\textcolor{pubcol!85!black}{$A_{\mathrm{pub}}(c_j)$}\\[0pt]
   \textcolor{privcol!60}{\rule{5pt}{5pt}}\,\textcolor{privcol!85!black}{$A_{\mathrm{priv}}(c_j)$}};

\node[mech] (P1) at (10.55,1.2)  {\textbf{SDG-CI}\\[-2pt]{\scriptsize or}\\[-2pt]\textbf{ANON-CI}};
\node[mech] (P2) at (10.55,-1.2) {\textbf{SDG-CI}\\[-2pt]{\scriptsize or}\\[-2pt]\textbf{ANON-CI}};
\draw[neuar] (T1g.east) -- (P1.west);
\draw[neuar] (T2g.east) -- (P2.west);

\drecq{d1}{10.55}{2.25}{neu}{neu}{neu}{neu}{privrec}
\drecq{d2}{10.55}{-2.25}{neu}{neu}{neu}{neu}{privrec}
\node[dlbl,anchor=east] at (d1.west) {$\mathcal{D}$};
\node[dlbl,anchor=east] at (d2.west) {$\mathcal{D}$};
\draw[inar] (d1.south) -- (P1.north);
\draw[inar] (d2.north) -- (P2.south);

\drecq{p1}{12.40}{1.80}{prot}{prot}{prot}{keep}{relrec}
\drecq{p2}{12.40}{1.20}{prot}{prot}{prot}{keep}{relrec}
\drecq{p3}{12.40}{0.60}{prot}{prot}{prot}{keep}{relrec}
\node[grp,draw=pubcol!45,fit=(p1)(p2)(p3),
      label={[text=pubcol!85!black,font=\small,inner sep=1pt]above:$\hat{\mathcal{D}}_1$}] (O1) {};
\drecq{q1}{12.40}{-0.60}{keep}{prot}{keep}{keep}{relrec}
\drecq{q2}{12.40}{-1.20}{keep}{prot}{keep}{keep}{relrec}
\drecq{q3}{12.40}{-1.80}{keep}{prot}{keep}{keep}{relrec}
\node[grp,draw=pubcol!45,fit=(q1)(q2)(q3),
      label={[text=pubcol!85!black,font=\small,inner sep=1pt]below:$\hat{\mathcal{D}}_2$}] (O2) {};
\draw[outar] (P1.east) -- (O1.west |- P1.east);
\draw[outar] (P2.east) -- (O2.west |- P2.east);

\pic (ua2) at (13.85,1.2)  {audience};
\pic (ub2) at (13.85,-1.2) {audience};
\node[audlbl,below=0pt of ua2-box] {$R_1$};
\node[audlbl,below=0pt of ub2-box] {$R_2$};
\draw[outar] (O1.east |- ua2-box.west) -- (ua2-box.west);
\draw[outar] (O2.east |- ub2-box.west) -- (ub2-box.west);


\draw[black!22,dashed] (4.88,-2.7) -- (4.88,2.7);
\draw[black!22,dashed] (9.42,-2.7) -- (9.42,2.7);

\node[stepnum] (n1) at (0.30,2.98) {1};
\node[lbl,anchor=west,text=black!65] at (n1.east) {Extract};
\node[stepnum] (n2) at (5.13,2.98) {2};
\node[lbl,anchor=west,text=black!65] at (n2.east) {Classify (per audience)};
\node[stepnum] (n3) at (9.67,2.98) {3};
\node[lbl,anchor=west,text=black!65] at (n3.east) {Privatize};

\end{tikzpicture}}
\caption{\small{Overview of Ext-Classify-Priv (private data in red, public and released data in green).
\textbf{(1)} Extract a schema with attributes $A=\{a_1,\dots,a_k\}$ from the private records $\mathcal{D}$.
\textbf{(2)} Each audience $R_j$ comes with its own release context $c_j$, which Classify uses to split $A$ into $A_{\mathrm{pub}}(c_j)$ and $A_{\mathrm{priv}}(c_j)$. The same corpus yields different splits: the advertiser $R_1$ gets three attributes protected, the public re-post $R_2$ only one.
\textbf{(3)} Privatize $\mathcal{D}$ with respect to $A_{\mathrm{priv}}(c_j)$ using SDG-CI or ANON-CI, and release $\hat{\mathcal{D}}_j$ to $R_j$. }}
\label{fig:pipeline}
\end{figure*}

Contextual Integrity (CI) \cite{nissenbaum2004privacy} offers a principled way to decide what each audience is allowed to see. CI holds that whether an information flow is appropriate depends on its context: who sends the information, who receives it, for what purpose, and the norms under which it was originally shared. Applied to a data release, CI tells us which attributes may flow to a given audience. However, CI alone does not privatize data. It is a normative framework that identifies \emph{which} information must not flow, but not \emph{how} to prevent it from flowing, or how much protection the release provides. The closest work, 1-2-3 Check \cite{check123}, uses a CI-guided LLM to flag information to remove, but it assumes a single audience and gives no privacy guarantee.

In this paper, we ask, \emph{given a private text corpus and several audiences with different needs and norms, can we automatically determine what to protect for each audience, and does audience-conditioned privatization yield more useful data than a single audience-agnostic release?}
To this end, we propose audience-conditioned privatization, which separates what to protect from how to protect it. CI determines which attributes are inappropriate for the intended audience, and a privatization mechanism protects only those attributes, preserving the rest. We realize this in \textsc{Ext-Classify-Priv}, a three-step pipeline (Figure~\ref{fig:pipeline}). \Ext{} derives a schema of attributes from the corpus and extracts each record's values using a locally hosted model, so raw text does not leave the data holder. \Classify{} labels each attribute as private or public for the intended audience, using an LLM prompted with the audience's CI context. \Priv{} then protects only the private attributes, using one of the two methods: (a)~\sdgci{} generates synthetic text with a formal Feature-DP \cite{fdp} guarantee, which bounds what the release reveals about a record beyond the attributes the audience is allowed to see, (b)~\anci{} rewrites each real record with an empirical privacy guarantee. Because the ``public'' attributes are not perturbed in these approaches, they are preserved with high fidelity. Our main contributions are:
\begin{itemize} [noitemsep, topsep=0pt]
\item \textbf{Audience-conditioned privatization framework} that uses CI to decide, per audience, which extracted attributes to protect.
\item \textbf{Two instantiations of \Priv{}:} \sdgci{}, with Feature-DP, and \anci{}.
\item \textbf{Audience-specific vs.\ single release.} On Reddit, audience-specific releases are more faithful for every audience (MAUVE 0.350 with MAPLE $\to$ 0.400--0.546 with \sdgci{} at $\varepsilon=1$).
\end{itemize}

\section{Related Work}\label{sec:rw_privatization}
Existing approaches to privatizing text fall into two families, summarized in Tab.~\ref{tab:related-work}. \textit{SDG methods} generate artificial data that is close in distribution to the private data without replicating individual records. \emph{Anonymization} directly removes sensitive content from the original text and leaves the rest intact.

\vspace{0.3em}
\noindent
\textbf{DP synthetic text.} DP synthetic text methods range from fine-tuning with DP-SGD~\cite{yu2024differentially} to inference-only Private Evolution (PE)~\cite{xiedifferentially, zou2026pesgd, hou2024pre}, which uses LLMs only through API calls. These methods satisfy standard record-level DP~\cite{dwork2014algorithmic} and protect every part of a record equally. SecPE~\cite{wang2026secret} relaxes this with secret-aware DP~\cite{ganesh2025hush}, but treats a whole record as secret if it contains predefined sensitive tokens, so it reduces to standard DP when every record does, as in our setting. \sdgci{} builds on PE, for its low cost and access to state-of-the-art API models, and protects at the level of attributes, spending budget only on those deemed sensitive for the intended audience.

\vspace{0.3em}
\noindent
\textbf{Anonymization.} LLM-based anonymizers~\cite{staab2025language, rupta} iteratively infer and redact private information, and DP rewriting~\cite{mattern-etal-2022-limits} paraphrases text under local DP. Both protect a fixed, predefined set of attributes or the whole record, which hinders real-world deployment~\cite{ochs2026conundrum, mattern-etal-2022-limits}. 1-2-3 Check~\cite{check123} derives what to remove using CI, but assumes a single audience. \anci{} builds on RUPTA~\cite{rupta} and derives private attributes per audience with \Ext{} and \Classify{}.

\vspace{0.3em}
\noindent
\textbf{Contextual integrity.} CI has been used to evaluate LLMs' privacy reasoning~\cite{mireshghallah2024can} and to limit what LLM agents disclose~\cite{bagdasarian2024airgapagent}, and has been related to DP in theory~\cite{comb_dp_ci}. None of these works privatize a corpus for release to different audiences. We use CI only to decide \emph{what} to protect for each audience, and leave \emph{how} to the privatization mechanism.

\begin{table}[t]
\centering
\footnotesize
\setlength{\tabcolsep}{3pt}
\begin{tabularx}{\columnwidth}{@{}>{\raggedright\arraybackslash\leftskip=0.6em\hangindent=0.6em}X ccc@{}}
\toprule
\multicolumn{1}{@{}l}{\textbf{Approach}} & \textbf{Aud.} & \textbf{Unit} & \textbf{Guarantee} \\
\midrule
\multicolumn{4}{@{}l}{\textit{Synthetic data generation}} \\
DP fine-tuning~\cite{yu2024differentially, hu2025actg}     & \xmark & Record & DP \\
Inference DP~\cite{xiedifferentially, chien2026maple} & \xmark & Record & DP \\
Secret-aware DP~\cite{wang2026secret}                         & \xmark & Secret & DP \\
\textbf{\textsc{SDG-CI} (ours)}                            & \cmark & Attr.  & Feat.-DP \\
\midrule
\multicolumn{4}{@{}l}{\textit{Anonymization}} \\
DP rewriting~\cite{mattern-etal-2022-limits}               & \xmark & Record & Local DP \\
LLM-based~\cite{staab2025language, rupta}                  & \xmark & Attr.  & Empirical \\
CI-guided~\cite{check123}                                  & \pmark & Attr.  & None \\
\textbf{\textsc{ANON-CI} (ours)}                           & \cmark & Attr.  & Empirical \\
\bottomrule
\end{tabularx}
\caption{\small{Text privatization approaches. In both categories, prior work either ignores the audience (\xmark) or assumes a single fixed one (\pmark); we add audience conditioning to each.
\textbf{Aud.}: the protected attributes depend on who receives the data.
\textbf{Unit}/\textbf{Guarantee}: what is protected and how.}}
\label{tab:related-work}
\end{table}

\section{Preliminaries}\label{sec:prelim}
\textbf{Problem formulation.} A data holder maintains a private corpus $D=(t_1,\ldots,t_n)$ and wishes to release privatized versions of it to audiences $R_1,\ldots,R_m$. Each audience $R_j$ comes with a release context $c_j$ (defined below), which includes its use case $u_j$. Each audience receives only its own release (i.e.,~non-colluding audiences) and we analyze each release separately. We describe the information in a text through \emph{attributes} (e.g., \textit{age}, \textit{topic}), given by a set $\A$ of attribute names, each with a finite domain $\X_a$, which together form a schema $\mathcal S=(\A,\{\X_a\}_{a\in\A})$. A text $t$ has attribute values $x\in\X=\prod_{a\in\A}\X_a$, $x[B]$ denotes the values on $B\subseteq\A$, and we call $z=(t,x)$ a \emph{record}. Neither party knows the schema in advance, since an audience knows its use case but not what the corpus contains, and the data holder has no predefined list of sensitive attributes. The goal is to release to each $R_j$ a corpus $\widehat{D}_j$ that is useful for $u_j$ while protecting what $R_j$ should not learn, without either party sharing the attributes in advance.

\vspace{0.3em}
\noindent
\textbf{Contextual integrity.} CI deems an information flow appropriate if it conforms to the norms of the context in which the information was originally shared \cite{nissenbaum2004privacy}. A flow is described by its data subject, sender, recipient, information type, and transmission principle, and we also make the use case explicit, following \cite{11369787}. In our setting, individuals first share their text with the data holder, and this original context sets the norms. A release then provides information about these individuals from the data holder to $R_j$ for use case $u_j$. We call $c_j=(R_j,u_j)$ the \emph{release context}, and hold the transmission principle fixed across audiences, since data is always released under the data owner's sharing terms. Since the data subjects and sender are also fixed, the only remaining flow parameter is the information type, which in our setting is an attribute $a \in \A$.

Each release flow is therefore determined by $c_j$ and an attribute $a\in\A$, and the release context partitions attributes into public attributes $\Apub(c_j)=\{a\in\A \,|\, (c_j,a)\text{ appropriate}\}$
and private attributes $\Apriv(c_j)=\A\setminus\Apub(c_j)$. An attribute is public not because it is harmless in general, but because this audience may learn it under the norms of the original context, e.g., house purchasing price for an internal fairness audit but not for an advertiser. An \emph{audience-conditioned} mechanism $\mathcal M(D;c_j)$ protects $\Apriv(c_j)$ and preserves $\Apub(c_j)$. Audience-agnostic SDG methods always set $\Apriv(c_j)=\A$.

\vspace{0.3em}
\noindent
\textbf{Feature-DP.} Feature-DP \cite{fdp} provides a formal guarantee for such a release. Under context $c_j$, the disclosed part of a record is $\Psi_j(z)=x[\Apub(c_j)]$.\footnote{We take the privacy unit to be one text.} A mechanism $\mathcal M$ is $(\varepsilon,\delta)$-Feature-DP relative to $\Psi_j$ if there is a simulator $\mathrm{sim}$ such that for every corpus $D$, record $z$, and event $E$, with $Y=\mathcal M(D\cup\{z\})$ and $Z=\mathrm{sim}(D,\Psi_j(z))$, we have $\Pr[Y\in E]\le e^{\varepsilon}\Pr[Z\in E]+\delta$, and symmetrically with $Y,Z$ swapped. The simulator sees the other records and $\Psi_j(z)$, but neither $x[\Apriv(c_j)]$ nor the text $t$. The release therefore reveals little about a record beyond what $R_j$ may learn under CI. With $\Apub(c_j)=\emptyset$, this is standard DP under add/remove adjacency.

\section{Methods}\label{sec:methods}
\textsc{Ext-Classify-Priv} consists of three steps (Figure~\ref{fig:pipeline}). \Ext{} runs once per corpus, while \Classify{} and \Priv{} run once per audience.

\subsection{Extract}
\label{sec:extract}
CI reasons about information types, which a text corpus usually does not come with, since neither the audience nor the data holder knows in advance which attributes it contains (Sec.~\ref{sec:prelim}). \Ext{} makes them explicit. It first derives the categorical schema $\mathcal S$ by prompting a locally hosted LLM to propose attributes and their possible values (App.~\ref{app:schema_deriving}), and coarsens attributes with large domains to categorical features as needed (App.~\ref{app:coarsening}). It then applies record-wise extraction map $\Ext$, run on the same local LLM, to each text, giving its attribute values $x_i=\Ext(t_i)$ and the Table $T=\{x_i\}_{i=1}^n$. 
The extracted attributes are what \Classify{} judges and what both \Priv{} methods protect or preserve. 
The schema itself is also a valuable artifact to share, for quickly analyzing new datasets whenever in-depth human review of the dataset is impossible or infeasible. As in tabular SDG~\cite{mckenna_aim}, we treat $\mathcal S$ as public and do not charge its derivation to the privacy budget.\footnote{Formally, this requires $\mathcal S$ not to depend on the private values in $D$, e.g., a schema prompted from a dataset description or derived with DP (App.~\ref{app:schema_deriving}).} In practice, the schema should be verified by a human before use.

\subsection{Classify}
Given a release context $c_j$, \Classify{} computes the partition $(\Apub(c_j),\Apriv(c_j))$ of Section~\ref{sec:prelim}. We prompt an LLM with $c_j$ and the attribute names in $\mathcal S$ (prompt in App.~\ref{app:prompts}), and it marks an attribute private if disclosing it to $R_j$ would violate the norms of the original context. \Classify{} sees only $\mathcal S$ and $c_j$, never the records. For instance, for Reddit posts, \textit{relationship status} may be private when the data is sent to advertisers but public when it is re-posted to r/relationship\_advice. The LLM's output is a proposal. In sensitive domains, a human expert, such as a privacy or legal reviewer, verifies the partition before any data is released. For our proprietary RealEstate data, a legal expert reviewed the partition for 3 audiences (App.~\ref{app:expert_ci_review}).

\subsection{Privatize}
Finally, \Priv{} protects $\Apriv(c_j)$ to generate a privatized corpus $\widehat{D}_j$ using one of two methods. \sdgci{} (Alg.~\ref{alg:sdg_ci}) produces synthetic text with a formal Feature-DP guarantee. \anci{} (Alg.~\ref{alg:rupta_ci}) rewrites each real record, providing an empirical privacy guarantee. Models and hyperparameters are listed in App.~\ref{app:hyperparameters}.

\vspace{0.3em}
\noindent
\textbf{SDG Approach.}
\sdgci{} builds on MAPLE \cite{chien2026maple} by combining tabular synthesis via AIM \cite{mckenna_aim} with Private Evolution (PE) \cite{xiedifferentially} to separate \emph{what} a synthetic record says from \emph{how} it is said. Both steps are necessary. Prompting an LLM directly with a synthetic row omits the tone, phrasing, and structure of real records, which PE recovers by iteratively filtering candidates against real data. Conversely, PE without tabular rows converges slowly and produces low-diversity text \cite{chien2026maple}. Whereas MAPLE treats all attributes as private, \sdgci{} spends privacy budget only on $\Apriv(c_j)$.

\vspace{0.3em}
\noindent
\textbf{Step 1: Feature-DP tabular generation.} We modify AIM \cite{mckenna_aim}, which provides standard record-level DP, to model the conditional distribution $P(\Apriv(c_j) \mid \Apub(c_j))$ (Algorithm~\ref{alg:cond_aim}), thus supporting Feature-DP guarantees relative to $\Psi_j$. Because the simulator is given $\Psi_j(z)$, statistics that depend only on public attributes require no noise. We therefore drop marginals over $B\subseteq\Apub(c_j)$, and the exponential mechanism selects only from a workload $W$ of marginals (of order at most three) containing at least one private attribute. This reduces the noise per measurement at a given $\varepsilon$.
During synthesis, public attributes are copied from real records, and private attributes are sampled sequentially, each conditioned on $\Apub(c_j)$ and the previously drawn private attributes.

\vspace{0.3em}
\noindent
\textbf{Step 2: PE text generation.} We keep MAPLE's PE stage where an LLM is prompted with synthetic rows to generate texts. In each PE iteration, each private record votes for its nearest synthetic candidate by cosine similarity, and Gaussian noise is added to the resulting vote histogram. The candidates with most votes are retained as seeds for the next round of variations (Figure~\ref{fig:pipeline}). The LLM never sees private text, which enters only through the noisy votes. Because each record votes with its full text, this stage protects everything in the text beyond $\Psi_j(z)$, including content that the schema does not capture.

\vspace{0.3em}
\noindent
\textbf{Privacy guarantee.} Both steps are accounted for in $\rho$-zCDP, following AIM. With $\rho=\rho_{\mathrm{AIM}}+\rho_{\mathrm{PE}}$, \sdgci{} satisfies $(\varepsilon,\delta)$-Feature-DP relative to $\Psi_j$, where $(\varepsilon,\delta)$ is the standard conversion of $\rho$ (App.~\ref{app:dp_proof}). 

\vspace{0.3em}
\noindent
\textbf{Anonymization Approach.}
\anci{} builds on RUPTA \cite{rupta}, which iteratively rewrites each record until an LLM privacy evaluator can no longer infer a predefined set of private attributes, while an LLM utility evaluator checks that a task label remains recoverable (details in Algorithm~\ref{alg:rupta_ci}). We make two modifications. First, the private attributes are $\Apriv(c_j)$, derived from the audience's context rather than predefined, so the same record is rewritten differently for each audience. Second, RUPTA's utility evaluator checks only the task label, so a rewrite could erase public attributes unrelated to it. We extend the utility check to require that $\Apub(c_j)$ also remain recoverable from $\widehat{D}_j$, which preserves exactly what the audience is allowed to see.

\input{fig_spider_results_concise}

\section{Experimental Setup}
For a fair comparison, all the baseline methods receive the same $S$, described in Tab.~\ref{tab:attributes}, and all methods share the same backbone LLM, Llama-3.1-8B-Instruct \cite{grattafiori2024llama}, unless indicated otherwise (with \Classify{} using Gemma-3-27B \cite{gemma}, Tab.~\ref{tab:hyperparams}). All results are from a single run.

\vspace{0.3em}
\noindent
\textbf{Datasets.} We evaluate our method on the synthetic Reddit dataset \cite{staab2025language, rupta}, privatizing it for different realistic audiences. It consists of Reddit question-and-answer pairs alongside human-annotated personal attributes of the writer of the answer. We also evaluate on the bioRxiv dataset \cite{chien2026maple, xiedifferentially}, and extend the evaluation to our own proprietary real estate (RealEstate) conversational set, consisting of 1,000 conversations between homebuyers and real estate agents. To validate the partition, a legal expert reviewed the split on our proprietary dataset for three realistic audiences (App.~\ref{app:expert_ci_review}). We report results for the Internal Audit audience.

\vspace{0.3em}
\noindent
\textbf{Baselines.} We compare \sdgci{} against {Aug-PE}~\cite{xiedifferentially} and {MAPLE}~\cite{chien2026maple}. For each experiment, we set the number of PE iterations to 2 for MAPLE and \sdgci{}, and 9 for Aug-PE, following the best results captured in previous research \cite{chien2026maple}.
\footnote{Our $\varepsilon$ covers only the private attributes, so it is not directly comparable to the baselines' record-level $\varepsilon$. Matching it shows how much fidelity is gained by not protecting public attributes, which is precisely the setting our method is designed for.}

 We compare \anci{} against {RUPTA}~\cite{rupta} and
{1-2-3 Check}~\cite{check123}. Both baselines protect a fixed set of private attributes, the default partition in Tab.~\ref{tab:audiences_datasets}, for every audience.

\paragraph{Evaluation Metrics.}
\label{sec:eval_metrics} We have 4 categories of metrics.

\noindent
\textit{Privacy.} We run an attribute inference attack (AIA), measuring how well an adversary can infer private attributes from the release $\hat{D}$, for both SDG \cite{zhao2021feasibility} and anonymization methods \cite{staab2025language}. For more details, see App.~\ref{app:metrics}. 
 
\noindent
\textit{Fidelity.} We assess the distributional similarity between the synthetic data and the real corpus using 3 metrics:
(a) {MAUVE} \cite{pillutla2021mauve}, which measures the  distributional distance between the synthetic and real data via divergence frontier; (b) {Per-attribute Jensen-Shannon Divergence (JSD)} \cite{oks} on the attributes extracted from the private text \cite{chien2026maple}; (c) {KL} and {TV} divergences on real and synthetic embeddings \cite{chung1989measures, xiedifferentially}.
 
\noindent
\textit{Utility.} We evaluate the utility of the synthetic data in a Train-on-Synthetic, Test-on-Real (TSTR) setup. We fine-tune a RoBERTa-base classifier \cite{liu2019roberta} on the synthetic texts to predict the dataset's utility label (Tab.~\ref{tab:audiences_datasets}) and report accuracy on a held-out set of real records \cite{xiedifferentially}. Since this label is public, our methods preserve it without noise. This task is a realistic proxy for downstream analyses, such as those on our proprietary data. 
 
\noindent
\textit{Fairness.} Following \cite{ganev2022robin}, we measure subgroup fairness
by fine-tuning a RoBERTa-base \cite{liu2019roberta} classifier on $\widehat{D}_j$ to predict the topic, testing it on held-out real records, and computing the absolute accuracy gap between the performance on texts associated with female and male authors, $|\Delta| = |\mathrm{acc}_{\text{female}} -\mathrm{acc}_{\text{male}}|$.\footnote{We only include the results on Reddit, as the only dataset with demographic attributes.}

\vspace{0.4em}
\noindent
\textbf{Evaluation Configurations.} Tab.~\ref{tab:audiences_datasets} lists the utility labels, release contexts, and  $\Apub(c_j)$ and $\Apriv(c_j)$ for each dataset. These default partitions are used in every experiment in the paper except Figure~\ref{fig:spider_audiences}(a-c) which follows the audiences in Tab.~\ref{tab:audiences_reddit}, where we vary the downstream audience to obtain different attribute splits and compare the output quality. 

\section{Results}
\label{sec:results}
%
Fig.~\ref{fig:spider_datasets} shows the performance of our method on public datasets compared to baselines. Our privatized data tends to be better aligned with the real data due to Feature-DP properties, which permit us to preserve public attributes as is. According to our experiments in App.~\ref{app:epsilon_results}, our method tends to have better fidelity than the baselines. The advantage of \textsc{SDG-CI} is lower at higher $\varepsilon$, which is to be expected, since at high privacy budgets there is less relative advantage to releasing the public attributes without noise. Fig.~\ref{fig:spider_audiences}(a--c) shows that we are able to get better performance for each audience when compared to single release.
Results on our proprietary dataset are in App.~\ref{app:realestate} (Figure~\ref{fig:spider_realestate}), where we obtain substantially higher MAUVE despite our JSD not being better (see Tab.~\ref{tab:spider_realestate} for details).

\vspace{0.3em}
\noindent
\textbf{Choosing the Right Method.}
The benefit of audience conditioning grows with the number of attributes the audience is allowed to see. By definition (Sec.~\ref{sec:prelim}), these are the attributes whose flow to the audience is appropriate. The largest gains therefore come from audiences close to the data's original context, such as a community where users already self-disclose information, or an internal team like the RealEstate Internal Audit team. On Reddit, SDG-CI's MAUVE gain over MAPLE is $+0.196$ for the re-post audience (8 public attributes), $+0.129$ for the ML vendor (6), and $+0.050$ for the advertiser (2). For audiences that should learn almost nothing, such as advertisers building behavioral profiles, \textsc{Classify} marks nearly every attribute private and SDG-CI behaves much like MAPLE. For such audiences, an audience-agnostic release costs the data owner little, whereas for audiences close to the original context, common in real-world settings, audience conditioning yields more valuable insight from the same data. 

\vspace{0.4em}
\noindent
\section{Conclusion}
Our work introduces a new framework designed to fill in the existing research gap in audience-conditioned privatization methods. It leads to better fidelity by taking into consideration the expected downstream audience of the privatized dataset.

\section*{Limitations}
Because the public attributes are released for every record, the public attributes can potentially identify an individual on their own, independent of the sensitive attributes the mechanism protects. Moreover, Feature-DP does not prevent inference \emph{from} public attributes, so a public attribute correlated with a private one (e.g., \textit{income\_level} and \textit{income}) partly reveals it, and a sensitive attribute mislabeled as public is released unprotected. This highlights the importance of choosing public and private attributes well, which is why we recommend human review of the split. It also highlights that our privacy guarantee constrains attribute inference rather than membership inference. This is an acknowledged limitation of Feature-DP \cite{fdp}. Deployments where the presence of a record in the training dataset itself is sensitive require standard DP rather than Feature-DP.
Each guarantee holds for a single release to its intended audience. If audiences share their releases or collude, they learn the attributes that were public for any of them, and repeated releases from the same corpus accumulate privacy loss on attributes that remain private. Our guarantee also covers coarsened attribute values, and mapping them back to specific values observed in the data, as in both MAPLE and \sdgci{}, falls outside it. We take one text as the privacy unit, so individuals who contribute several texts receive weaker protection. \anci{} provides only an empirical guarantee, and our results come from a single run.

\section*{Ethics Statement}
This work aims to let organizations use sensitive user text while protecting the individuals who contributed it. The Reddit dataset consists of synthetic comments with synthetic personal attributes \cite{staab2025language}, and bioRxiv is publicly available. Extraction and privatization run on locally hosted models, so raw text did not leave the data holder, and neither the data nor its privatized versions are released.
\Classify{} relies on an LLM, which may mislabel a sensitive attribute as public. We therefore treat its output as a proposal that a human expert should verify before release, as we did for RealEstate. Public attributes are released exactly and may enable re-identification in combination, and \anci{} provides only an empirical guarantee. Practitioners should choose the method and the attribute split according to the sensitivity of the data and the trust placed in each audience.

\bibliographystyle{plain}
\bibliography{references}
\appendix
 
 
\section{Deriving the Schema}
\label{app:schema_deriving}
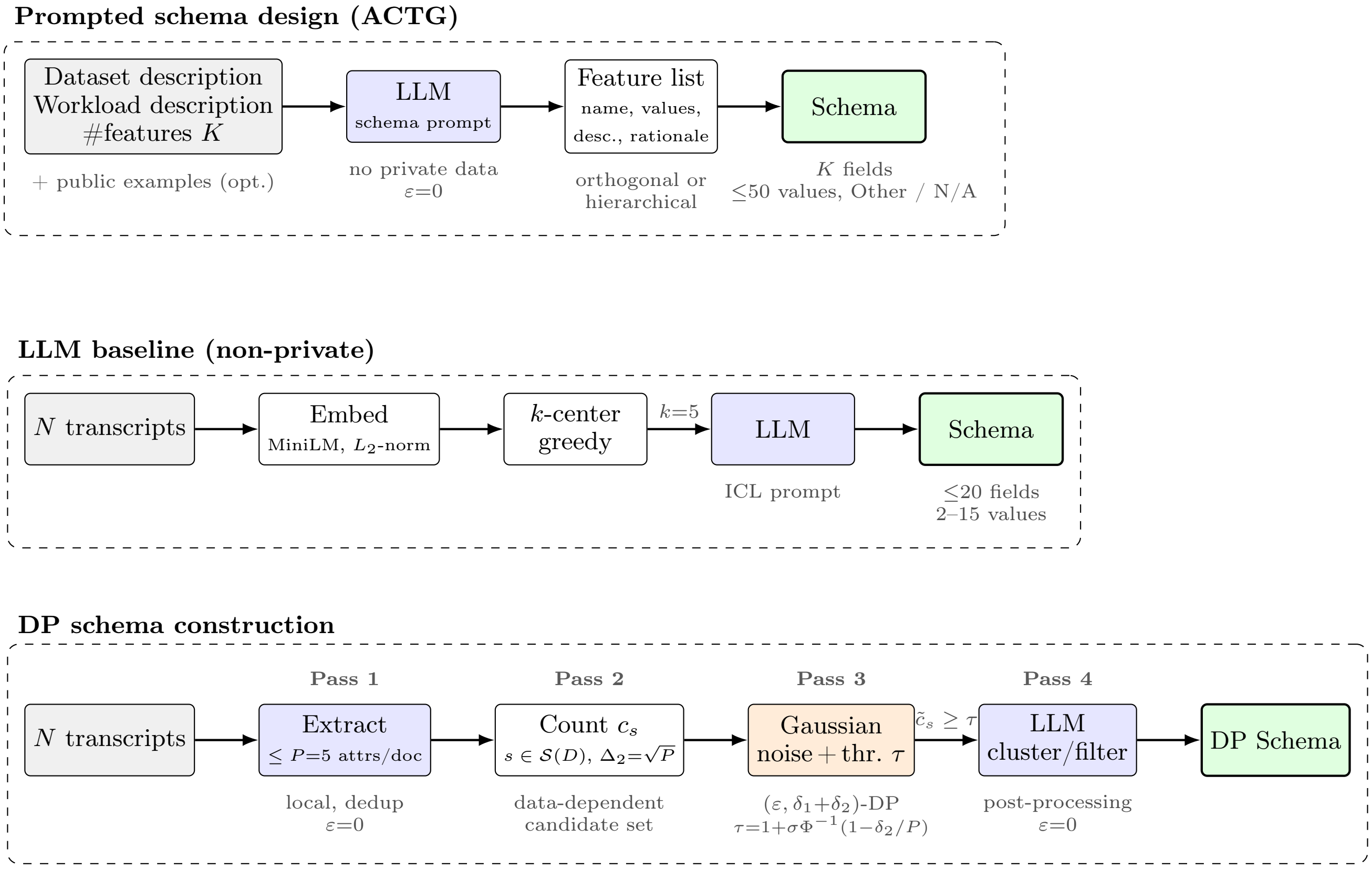
\begin{figure*}[t]
\centering
\begin{tikzpicture}[
  font=\small,
  node distance=6mm and 8mm,
  box/.style={draw, rounded corners=2pt, align=center, minimum height=9mm, minimum width=18mm, inner sep=3pt, fill=white},
  data/.style={box, fill=gray!12},
  llm/.style={box, fill=blue!10},
  priv/.style={box, fill=orange!15},
  result/.style={box, fill=green!12, thick},
  pass/.style={font=\scriptsize\bfseries, gray!70!black},
  note/.style={font=\scriptsize, text=gray!60!black, align=center},
  arr/.style={-{Latex[length=2mm]}, thick},
  panel/.style={draw, dashed, rounded corners=4pt, inner sep=6pt},
  title/.style={font=\small\bfseries, anchor=south west}
]

\node[data] (t3) {Dataset description\\Workload description\\\#features $K$};
\node[llm, right=of t3] (llm3) {LLM\\{\tiny schema prompt}};
\node[box, right=of llm3] (fl) {Feature list\\{\tiny name, values,}\\{\tiny desc., rationale}};
\node[result, right=of fl] (s3) {Schema};

\draw[arr] (t3) -- (llm3);
\draw[arr] (llm3) -- (fl);
\draw[arr] (fl) -- (s3);

\node[note, below=1mm of t3] (n5) {+ public examples (opt.)};
\node[note, below=1mm of llm3] (n6) {no private data\\$\varepsilon{=}0$};
\node[note, below=1mm of fl] (n6b) {orthogonal or\\hierarchical};
\node[note, below=1mm of s3] (n7) {$K$ fields\\$\le$50 values, Other / N/A};

\begin{scope}[on background layer]
  \node[panel, fit=(t3)(s3)(n5)(n6)(n6b)(n7)] (P3) {};
\end{scope}
\node[title] at (P3.north west) {Prompted schema design (ACTG)};

\node[data, anchor=north west] (t1) at ([yshift=-30mm]t3.south west) {$N$ transcripts};
\node[box, right=of t1] (emb) {Embed\\{\tiny MiniLM, $L_2$-norm}};
\node[box, right=of emb] (kc) {$k$-center\\greedy};
\node[llm, right=of kc] (llm1) {LLM};
\node[result, right=of llm1] (s1) {Schema};

\draw[arr] (t1) -- (emb);
\draw[arr] (emb) -- (kc);
\draw[arr] (kc) -- node[above, note] {$k{=}5$} (llm1);
\draw[arr] (llm1) -- (s1);
\node[note, below=1mm of llm1] (n1) {ICL prompt};
\node[note, below=1mm of s1] (n2) {$\le$20 fields\\2--15 values};

\begin{scope}[on background layer]
  \node[panel, fit=(t1)(s1)(n1)(n2)] (P1) {};
\end{scope}
\node[title] at (P1.north west) {LLM baseline (non-private)};

\node[data, anchor=north west] (t2) at ([yshift=-30mm]t1.south west) {$N$ transcripts};
\node[llm, right=of t2] (ex) {Extract\\{\tiny $\le P{=}5$ attrs/doc}};
\node[box, right=of ex] (agg) {Count $c_s$\\{\tiny $s\in\mathcal{S}(D)$, $\Delta_2{=}\sqrt{P}$}};
\node[priv, right=of agg] (gm) {Gaussian\\noise\,+\,thr.~$\tau$};
\node[llm, right=of gm] (cl) {LLM\\cluster/filter};
\node[result, right=of cl] (s2) {DP Schema};

\draw[arr] (t2) -- (ex);
\draw[arr] (ex) -- (agg);
\draw[arr] (agg) -- (gm);
\draw[arr] (gm) -- node[above, note] {$\tilde c_s \ge \tau$} (cl);
\draw[arr] (cl) -- (s2);

\node[pass, above=1mm of ex] (p1) {Pass 1};
\node[pass, above=1mm of agg] {Pass 2};
\node[pass, above=1mm of gm] {Pass 3};
\node[pass, above=1mm of cl] {Pass 4};

\node[note, below=1mm of ex] (n3) {local, dedup\\$\varepsilon{=}0$};
\node[note, below=1mm of agg] (n3b) {data-dependent\\candidate set};
\node[note, below=1mm of gm] (n4) {$(\varepsilon,\delta_1{+}\delta_2)$-DP\\{\tiny $\tau{=}1{+}\sigma\Phi^{-1}(1{-}\delta_2/P)$}};
\node[note, below=1mm of cl] (n4b) {post-processing\\$\varepsilon{=}0$};

\begin{scope}[on background layer]
  \node[panel, fit=(t2)(s2)(p1)(n3)(n3b)(n4)(n4b)] (P2) {};
\end{scope}
\node[title] at (P2.north west) {DP schema construction};

\end{tikzpicture}
\caption{Schema construction. Top: ACTG-style prompted schema design \cite{hu2025actg}, where the schema is written from \textbf{human knowledge} (dataset and workload descriptions) without private data; each private document is then annotated against the schema by a trusted oracle LLM. Middle: non-private LLM baseline. Bottom: four-pass DP schema construction; only Pass~3 consumes privacy budget, and the positive threshold $\tau$ guards against releasing strings that exist only because of a single transcript, since the candidate set $\mathcal{S}(D)$ is itself data-dependent.}
\label{fig:schema-pipeline}
\end{figure*}

We experiment with three different ways of producing attribute schema $\mathcal S$ when none is available. The methods are described below and in Figure \ref{fig:schema-pipeline}. In our experiments, we use the best schema as judged by humans. That means that for bioRxiv and Reddit, we use the schema provided by the original authors, and for RealEstate, we use a schema obtained by  using the local LLM approach and verifying it with humans. 
 
\begin{table*}[t]
\centering
\begin{tabular}{llrrrrr}
\toprule
Method & $\varepsilon$ & Fields & Signal$\uparrow$ & Balanced Entropy$\uparrow$ & Dominance$\downarrow$ & Coverage$\uparrow$ \\
\midrule
Hand-written & n/a & 47 & 0.144 & 0.33 & 0.72 & 2.19 \\
LLM & n/a & 20 & 0.440 & \textbf{0.96} & 0.58 & 3.14 \\
MAPLE & n/a & 8 & \textbf{0.718} & 0.65 & 0.59 & 2.84 \\
DP & $\infty$ & 20 & 0.278 & 0.69 & 0.57 & \textbf{3.24} \\
DP & 100 & 19 & 0.243 & 0.75 & \textbf{0.56} & 3.00 \\
DP & 10 & 15 & 0.236 & 0.63 & 0.65 & 2.65 \\
\bottomrule
\end{tabular}
\caption{Quality of the derived schemas on the RealEstate subset.}
\label{tab:schema_results}
\end{table*}
 
\subsection{Prompted Schema Design (ACTG)}
ACTG \cite{hu2025actg} and MAPLE \cite{chien2026maple} build the schema from human
domain knowledge. A templated prompt
(\cite{hu2025actg}, Fig.~6) requires human-authored
\textsc{dataset\_description}, a \textsc{workload\_description} describing the
downstream task the features should serve, and a target number of features
\textsc{num\_features}~$=K$. Optionally, a handful of publicly available or donated
examples from the general domain are appended as \textsc{\_formatted\_examples}.
 
The prompt instructs the LLM to propose $K$ categorical features that (i) mix
general-purpose attributes (e.g., formality, sentiment) with domain-specific ones
capturing the jargon, entities, and processes of the target data; (ii) are
orthogonal unless intentionally hierarchical, in which case a conditional feature
takes the value \emph{Not Applicable} when its parent condition is not met;
(iii) each have a fixed set of at most 50 explicitly enumerated values, with an
\emph{Other} category where appropriate; and (iv) avoid features that are trivial
or tailored to a single exemplar. The model returns a numbered list giving, for
each feature, its name, possible values, a description, and a rationale tied to
the workload. This list is used as schema $\mathcal S$.
 
Because no private transcript is read during schema design, this step consumes no
privacy budget ($\varepsilon = 0$). Feature extraction is then performed per
document against the fixed schema, either with a model run locally (e.g., Llama-3.1-8B-Instruct) or with a server-hosted oracle LLM
(e.g., gemini-2.5-flash-lite \cite{hu2025actg}). The trade-off relative to the two other methods for schema extraction is that ACTG requires the most domain knowledge. It assumes the author knows sufficient information about the dataset and workload description for every new domain to determine a comprehensive schema.
 
\subsection{LLM-Driven Construction}
The LLM baseline operates in two phases to produce an attribute schema.
 
First, we embed all $N$ transcripts using a pre-trained sentence transformer (\textsc{all-MiniLM-L6-v2}) and apply $L_2$-normalization. We then use $k$-center greedy selection strategy to select $k$ representative text samples. For our experiments, we set $k=5$.
 
In the second phase, we concatenate the $k$ selected transcripts and pass them as in-context-learning examples to a large language model. The prompt instructs the model to design a coding scheme describing $k$ passed examples, proposing up to 20 attributes, each with 2 to 15 categorical values.
 
\subsection{DP Schema Construction}
For DP Schema Construction, we follow previous work on DP domain discovery \cite{raman2026differentially} adjusting it to derive a set of valid \{attribute:value\} pairs to construct the schema in 4 passes.
 
\textbf{Pass 1: Attribute Extraction.}
For each of the $N$ transcripts, we prompt a LLM to extract up to $P=5$ attribute strings. Because this extraction is processed locally and independently per document, it consumes no privacy budget.
 
\textbf{Pass 2: Attribute Aggregation.}
We then compute the exact document frequency for every unique attribute proposed across the documents. Since each document can contribute a maximum of $P=5$ unique items, the $L_2$ sensitivity is $\Delta = \sqrt{5}$.
 
\textbf{Pass 3: Private Domain Discovery.}
We then apply the Gaussian mechanism to guarantee DP. For each unique string, we add independent Gaussian noise in proportion to the $L_2$ sensitivity $\Delta = \sqrt{5}$ and the target $(\varepsilon, \delta)$ privacy parameters (where $\delta = 1/N^2$). We then apply a threshold to filter out any string whose noisy count falls below threshold $\tau$.
 
\textbf{Pass 4: DP Release.}
The text strings that survive the thresholding step get added to our private schema as attributes and are clustered and filtered using an LLM. This step reads only the released strings, so by post-processing it costs no additional budget.
 
\subsection{Schema Extraction Results}
Table~\ref{tab:schema_results} compares the schemas produced by the pipelines in Figure~\ref{fig:schema-pipeline} on the subset of RealEstate data. Each schema is applied by the same LLM extractor to a random sample of 200 documents. \emph{Signal} is the fraction of documents in which a field receives a value other than ``not stated''. \emph{Balanced entropy} is the entropy of a field's value distribution divided by $\log_2$ of its number of distinct values. \emph{Dominance} is the share of the most common value. All three are averaged over fields. \emph{Coverage} is an LLM judge's 1--5 rating, on 100 subsampled documents, of how much of a document's content the extracted record captures, such that higher value means less information is lost. Without noise ($\varepsilon=\infty$) the DP pipeline gives the highest coverage, and as $\varepsilon$ decreases, both the number of fields and the coverage drop. We note that DP extraction of information from text is not a solved problem and even existing state-of-the-art solutions only work well at high privacy budgets \cite{annamalai2026cliopatra}.
 
\section{Coarsening of High-Cardinality Attributes}
\label{app:coarsening}
A marginal over $B\subseteq\A$ has $\prod_{a\in B}|\X_a|$ cells, and AIM adds noise of the same scale to every cell. Free-text attributes such as \emph{occupation} or \emph{city\_country} can take hundreds of values, so a three-way marginal involving two of them is infeasible to compute for large datasets. We therefore coarsen such attributes before the mechanism runs. When coarsening is enabled, an attribute with a domain of more than 10 values is coarsened into a set of predefined values constructed from the human-verified LLM output.
 
For each such attribute, an LLM (Llama-3.1-8B-Instruct) receives the attribute name and the list of values in $\X_a$, and maps each value to one of at most 5 buckets. For the Reddit dataset, for example, these values were used: 18--24, 25--34, 35--44, 45--54, 55+ for age, $<$20k, 20--49k, 50--79k, 80--119k, 120k+ for income, doctorate, master's, bachelor's, high school, other for education, etc. The result is a map $g_a:\X_a\to\mathcal B_a$ where $|\mathcal B_a|\le 5$.
 
Then for MAPLE and SDG-CI approaches, conditional AIM (Algorithm~\ref{alg:cond_aim}) is fitted on the coarsened table $\{g(x_i)\}$. The public attributes are then restored to their real values and for each private attribute, the coarsened value is replaced by a real value drawn at random from $g_a^{-1}(\text{bucket})$ (e.g.,\ ``chef'' rather than ``Food \& Hospitality'').
 
 
\section{Prompts}
\label{app:prompts}
The following is the prompt for \Classify{} step aimed at determining which attributes are sensitive for the given audience.
 
\begin{promptbox}[label=prompt:ci]{contextual-integrity privacy analyst}
\begin{promptverb}
You are a privacy analyst applying the theory of contextual
integrity to determine which information can and cannot be
disclosed from the sender to the recipient about the data
subject.
 
An information flow is appropriate only if it matches the norms
of the context in which the information was originally shared.
Judge each attribute below: if disclosing it about an
identifiable individual to this recipient, for this purpose,
would violate those norms, it must be privatized.
 
!slot[{context.describe()}]
  (e.g. Sender / data subject: Reddit users posting in public
   forums (attributes describe the post author).
   Recipient: an academic research team publishing aggregate
   findings on online discourse.
   Purpose: studying demographic inference from public
   social-media text.)
 
Attributes:
!slot[{attr_lines}]
 
For each attribute output an object with:
  "privatize": true or false
  "reason": one sentence naming the norm at stake
 
Output exactly one JSON object with keys !slot[{keys}] and no
other text.
\end{promptverb}
\end{promptbox}
 

\section{Algorithms}
\label{app:algos}
\subsection{\textsc{SDG-CI}}

Algorithm~\ref{alg:sdg_ci} summarizes \textsc{SDG-CI}. 

\textbf{Conditional AIM.} AIM \cite{mckenna_aim} is a marginal-based generator that follows the select-measure-estimate paradigm. Algorithm~\ref{alg:cond_aim} shows our modified version of AIM changed to support Feature-DP.

As in AIM's Algorithm~1 and in the MAPLE baseline, the fit begins by measuring the one-way marginal of every private attribute $a\in\Apriv(c_j)$ with Gaussian noise. These measurements are charged 10\% of $\rho_\text{AIM}$, with noise $\sigma_0=K/\sqrt{0.2\rho_\text{AIM}}$ on each of the $K=|\Apriv(c_j)|$ marginals (line~\ref{line:oneway}). Values whose noisy count is below $3\sigma_0$ are merged into a single bin per attribute. The remaining 90\% of $\rho_\text{AIM}$ is spent by the adaptive loop composed of three steps. The \emph{select} step (line~\ref{line:select}) uses the exponential mechanism \cite{mcsherry2007mechanism} to choose the candidate marginal with highest error measurement. The \emph{measure} step (line~\ref{line:measure}) releases that marginal with Gaussian noise \cite{dwork2014algorithmic}. The \emph{estimate} step reconciles the noisy measurements using Private-PGM \cite{McKenna_Private-PGM_2021}. Per-round budgets are annealed by doubling, and the last round is shrunk to the remainder, so the loop spends exactly $0.9\rho_\text{AIM}$.

We make three changes to AIM to accommodate Feature-DP. First, we restrict workload $W$ to only marginals of order at most three that involve at least one private attribute and have at most $10^7$ cells (line~\ref{line:workload}).
Second, we drop public-only candidates, so that no budget goes to measuring $p(\Apub(c_j))$, which the sampler never uses.
Third, private attributes are then drawn one at a time, conditioned on a subset $\mathrm{pa}_k$ of the record's public attributes and the previously drawn private attributes (lines~\ref{line:seq}--\ref{line:seq_draw}): $\hat x_i[a_k] \sim \hat{p}(a_k \mid \hat x_i[\mathrm{pa}_k])$. The subset is chosen without looking at the data, so that the conditional table has at most $10^7$ cells for efficiency.

\textbf{Private Evolution.} PE generates a candidate synthetic text based on the synthetic attributes produced by Conditional AIM. It is then iteratively voted on over $T_\text{PE}+1$ rounds. In each round, every private record votes for its nearest candidate in embedding space (line~\ref{line:vote}), and Gaussian noise is added to the vote histogram (line~\ref{line:noise}). The highest-voted candidates are retained (line~\ref{line:retain}) and varied by the LLM to form next iteration's population (line~\ref{line:variation}). 

\textbf{Privacy guarantee.} Both stages are calibrated under insertion/deletion adjacency, such that marginals and vote histograms have $\ell_2$ sensitivity 1, and the selection score $q_r$ has sensitivity $w_r$. In Appendix~\ref{app:dp_proof}, we show that \textsc{SDG-CI} satisfies $(\epsilon, \delta)$-Feature-DP relative to $\Psi_j$ for the total zCDP budget $\rho = \rho_\text{AIM} + \rho_\text{PE}$ split 0.1/0.9 following MAPLE \cite{chien2026maple} with the exception of schema $\mathcal S$ and the held-out in-context examples, which are assumed to be public, and the values of $x_i[\Apub(c_j)]$ which are released as is.

\begin{algorithm}[ht!]
\caption{Conditional AIM}
\label{alg:cond_aim}
\begin{algorithmic}[1]
\REQUIRE Table $T$, partition $(\Apub(c_j),\Apriv(c_j))$ with private attributes $a_1,\ldots,a_K$, privacy parameter $\rho_\text{AIM}$
\ENSURE Synthetic table $\widehat T$
\STATE \textbf{Hyper-parameters:} MAX-SIZE $=80$MB, $R=16|\A|$, $\alpha=0.9$
\STATE $W \leftarrow \{ r \subseteq \A : |r| \le 3,\ r \cap \Apriv(c_j) \neq \emptyset,\ \prod_{a\in r}|\X_a| \le 10^7 \}$ \label{line:workload}
\STATE $W^+ \leftarrow$ downward closure of $W$, without public-only sets;\; $w_r = \sum_{s \in W} |r \cap s|$ \label{line:candidates}
\STATE Measure the one-way marginals of $\Apriv(c_j)$ with noise $\sigma_0=K/\sqrt{0.2\rho_\text{AIM}}$, charging $0.1\rho_\text{AIM}$ \label{line:oneway}
\STATE Merge values with noisy count below $3\sigma_0$ \label{line:compress}
\STATE Initialize $\hat p_0$ with Private-PGM \label{line:init}
\STATE $\rho \leftarrow 0.9\rho_\text{AIM}$;\; $\sigma_1 \leftarrow \sqrt{R/(2\alpha\rho)}$;\; $\epsilon_1 \leftarrow \sqrt{8(1-\alpha)\rho/R}$;\; $\rho_\text{used} \leftarrow 0$;\; $t \leftarrow 0$
\WHILE{$\rho_\text{used} < \rho$}
    \STATE $t \leftarrow t + 1$;\; \textbf{if} $\rho-\rho_\text{used} < 2\bigl(\frac{1}{8}\epsilon_t^2 + \frac{1}{2\sigma_t^2}\bigr)$: set $\epsilon_t,\sigma_t$ so that this round spends exactly $\rho-\rho_\text{used}$ \label{line:final}
    \STATE $\rho_\text{used} \leftarrow \rho_\text{used} + \frac{1}{8}\epsilon_t^2 + \frac{1}{2\sigma_t^2}$
    \STATE select $r_t \in W^+$ (within MAX-SIZE) using the exponential mechanism with $q_r(T) = w_r\bigl(\|M_r(T) - M_r(\hat p_{t-1})\|_1 - \sqrt{2/\pi}\,\sigma_t\, n_r\bigr)$ \label{line:select}
    \STATE measure $\tilde y_t = M_{r_t}(T) + \mathcal N(0, \sigma_t^2 I)$ \label{line:measure}
    \STATE estimate $\hat p_t$ with Private-PGM from all measurements \label{line:estimate}
    \STATE anneal $\epsilon_{t+1}$ and $\sigma_{t+1}$: double the round's budget if $\|M_{r_t}(\hat p_t) - M_{r_t}(\hat p_{t-1})\|_2 \le \sqrt{2/\pi}\,\sigma_t\, n_{r_t}$
\ENDWHILE
\STATE copy $\hat x_i[\Apub(c_j)] \leftarrow x_i[\Apub(c_j)]$ for all $i$ \label{line:copy}
\FOR{$k = 1, \ldots, K$} \label{line:seq}
    \STATE sample $\hat x_i[a_k]$ from $\hat p_t(a_k \mid \hat x_i[\mathrm{pa}_k])$, with $\mathrm{pa}_k \subseteq \Apub(c_j) \cup \{a_1, \ldots, a_{k-1}\}$ \label{line:seq_draw}
\ENDFOR
\STATE expand merged values uniformly over the values merged into them \label{line:expand}
\RETURN $\widehat T$
\end{algorithmic}
\end{algorithm}

\begin{algorithm}[!h]
\caption{\textsc{SDG-CI}}
\label{alg:sdg_ci}
\begin{algorithmic}[1]
\REQUIRE Private corpus $D=(t_1,\ldots,t_n)$; schema $\mathcal S$; CI context $c_j$; privacy parameters $(\epsilon, \delta)$, $\delta=1/n$; number of PE iterations $T_\text{PE}$
\ENSURE Synthetic corpus $\widehat D_j$
\STATE $x_i\leftarrow\Ext(t_i)$ for all $i$, unless the attributes ship with the data;\; $T\leftarrow(x_1,\ldots,x_n)$ \hfill $\triangleright$ Appendix~\ref{app:schema_deriving} \label{line:sdg_extract}
\STATE $(\Apub(c_j),\Apriv(c_j))\leftarrow\Classify(\mathcal S,c_j)$ \hfill $\triangleright$ Appendix~\ref{app:prompts} \label{line:sdg_partition}
\STATE $\rho \leftarrow$ largest zCDP budget that converts to $(\epsilon, \delta)$ \hfill $\triangleright$ Appendix~\ref{app:dp_proof} \label{line:rho}
\STATE $\rho_\text{AIM} \leftarrow 0.1\rho$;\; $\rho_\text{PE} \leftarrow 0.9\rho$ \hfill \label{line:split}
\STATE $\widehat T\leftarrow\textsc{ConditionalAIM}\bigl(T,\Apub(c_j),$\newline
\hspace*{1.5em}$\Apriv(c_j),\rho_\text{AIM}\bigr)$ \hfill $\triangleright$ Algorithm~\ref{alg:cond_aim}
\STATE $\sigma_\text{PE} \leftarrow \sqrt{(T_\text{PE}+1) / (2\rho_\text{PE})}$ \label{line:pe_sigma}
\STATE $Z \leftarrow$ candidate texts generated by an LLM conditioned on each $\hat x\in\widehat T$ and held-out in-context examples \label{line:pe_init}
\FOR{$t = 0, \dots, T_\text{PE}$}
    \STATE \textbf{if} $t>0$: $Z \leftarrow Z \cup$ LLM-generated variations of the candidates in $Z$ \label{line:variation}
    \STATE Embed $Z$ and $D$; each $t_i$ votes for its nearest candidate in $Z$ with its utility label by $\ell_2$ distance, giving histogram $h$ \label{line:vote}
    \STATE $\tilde{h} \leftarrow h + \mathcal{N}(0, \sigma_\text{PE}^2 I)$ \label{line:noise}
    \STATE $Z \leftarrow$ highest-voted candidates under $\tilde{h}$, with per-label quotas computed from $\widehat T$ \label{line:retain}
\ENDFOR
\RETURN $\widehat D_j\leftarrow Z$ \label{line:pe_return}
\end{algorithmic}
\end{algorithm}

\subsection{\textsc{ANON-CI}}

Algorithm~\ref{alg:rupta_ci} summarizes \textsc{ANON-CI}. It shares extraction and classification steps with \textsc{SDG-CI} (lines~\ref{line:rupta_extract}--\ref{line:rupta_partition}). It also requires a utility label $y_i$ for each record, i.e., the record's label for the downstream task, such as the topic label used in our utility evaluation. Each record is then rewritten over up to $T_{\max}$ rounds. In each round, the rewriter generalizes the spans of the current rewrite that reveal private attributes (line~\ref{line:rewrite}). The privacy evaluator tries to infer each private attribute with $p_{\text{threshold}}$ guesses, and any attribute whose extracted value is among the guesses counts as leaked (line~\ref{line:privacy}). The utility evaluator checks that both $y_i$ and the values of $x_i[\Apub(c_j)]$ remain inferable (line~\ref{line:utility}). Checking $\Apub(c_j)$ is our extension, the original RUPTA checks only the label. The loop stops once no attribute leaks and all utility checks pass. Otherwise, the leaked attributes, the spans that revealed them, and any failed utility checks are returned to the rewriter as feedback (line~\ref{line:feedback}). Finally, each record is released as the privacy-safe rewrite that best preserves utility. If no round is privacy-safe, the rewrite with the fewest leaked attributes is released (line~\ref{line:release}).

\begin{algorithm}[!h]
\caption{\textsc{ANON-CI}}
\label{alg:rupta_ci}
\begin{algorithmic}[1]
\REQUIRE Private corpus $D=(t_1,\ldots,t_n)$ with utility labels $\{y_i\}_{i=1}^{n}$; CI context $c_j$; guess budget $p_{\text{threshold}}$; maximum rounds $T_{\max}$
\ENSURE Anonymized corpus $\widehat D_j$
\STATE $x_i\leftarrow\Ext(t_i)$ for all $i$ \hfill $\triangleright$ Appendix~\ref{app:schema_deriving} \label{line:rupta_extract}
\STATE $(\Apub(c_j),\Apriv(c_j))\leftarrow\Classify(\mathcal S,c_j)$ \hfill $\triangleright$ Appendix~\ref{app:prompts} \label{line:rupta_partition}
\FOR{$i = 1, \dots, n$}
    \STATE $\hat t_i^{(0)}\leftarrow t_i$;\; $f \leftarrow \emptyset$
    \FOR{$t = 1, \dots, T_{\max}$}
        \STATE \textbf{Rewrite:} $\hat t_i^{(t)}$ $\leftarrow$ rewriter LLM generalizes the spans of $\hat t_i^{(t-1)}$ that reveal $\Apriv(c_j)$, using feedback $f$ and preserving $\Apub(c_j)$ \label{line:rewrite}
        \STATE \textbf{Privacy:} for each $a \in \Apriv(c_j)$, the privacy evaluator infers $a$ from $\hat t_i^{(t)}$ with $p_{\text{threshold}}$ guesses; $L \leftarrow \{a : x_i[a] \text{ is among the guesses}\}$ \label{line:privacy}
        \STATE \textbf{Utility:} the utility evaluator checks whether $y_i$ and each value in $x_i[\Apub(c_j)]$ are inferable from $\hat t_i^{(t)}$; $U \leftarrow$ the items that are not \label{line:utility}
        \IF{$L = \emptyset$ \AND $U = \emptyset$}
            \STATE \textbf{break}
        \ENDIF
        \STATE $f \leftarrow (L, U)$, together with the spans that revealed $L$ \label{line:feedback}
    \ENDFOR
    \STATE $\hat t_i\leftarrow$ among the rounds with $L = \emptyset$, the rewrite that best preserves utility; if there is none, the rewrite with the smallest $|L|$ \label{line:release}
\ENDFOR
\RETURN $\widehat D_j=(\hat t_1,\ldots,\hat t_n)$
\end{algorithmic}
\end{algorithm}

 
\section{Proof of Feature Differential Privacy}
\label{app:dp_proof}

A record is $z=(t,x)$ with $x=\Ext(t)$, computed once per text, or, on Reddit and bioRxiv, the attributes shipped with the data, and its disclosed part under context $c_j$ is $\Psi_j(z)=x[\Apub(c_j)]$. Everything else about $z$, namely its private attributes $x[\Apriv(c_j)]$ and its text $t$, is protected. We assume that (A1) the guarantee is conditional on the schema $\mathcal S$, including its value lists and coarsening maps, on $\Ext$, and on the hyperparameters (Tab.~\ref{tab:hyperparams}), which we treat as fixed, such that it holds end to end only if they are chosen without access to $D$ or derived with DP (App.~\ref{app:schema_deriving}), whose cost then composes with $\rho$; (A2) the partition of the attributes into public and private ones, $(\Apub(c_j),\Apriv(c_j))$, depends only on $\mathcal S$ and $c_j$, not on the records in $D$; (A3) the LLM calls in Algorithm~\ref{alg:sdg_ci} see only rows of $\widehat T$ (real public values and synthetic private values), earlier candidates, and in-context examples from a split disjoint from $D$, which we treat as public (as in MAPLE, they are assumed donated) and which the guarantee does not protect; (A4) each synthetic bin is mapped back to a specific value drawn uniformly at random from a per-bin value set $\mathcal V$ (App.~\ref{app:coarsening}); and (A5) the LLMs and the embedder are pre-trained and not trained on $D$. In our experiments, $\mathcal V$ in (A4) is the set of distinct values that records in $D$ hold in each bin. Because this step reads $D$ and spends no privacy budget, it is not covered by the guarantee. Formally, the guarantee treats $\mathcal V$ as given, that is, unchanged when a record is added or removed, as with $\mathcal S$ in (A1), and what $\mathcal V$ reveals, namely which values occur in $D$, is not protected. 

Both stages of \sdgci{} are accounted in zCDP~\cite{bun2016concentrated}, so we state Feature-DP in zCDP form, where $\mathrm{D}_\alpha$ denotes the R\'enyi divergence of order $\alpha$.

\begin{definition}[Feature-zCDP]
\label{def:fzcdp}
A mechanism $\mathcal M$ is $\rho$-feature-zCDP relative to $\Psi_j$ if there is a simulator $\mathrm{sim}$ such that for every corpus $D$, record $z$, and $\alpha>1$, with $Y=\mathcal M(D\cup\{z\})$ and $Z=\mathrm{sim}(D,\Psi_j(z))$, we have $\mathrm{D}_\alpha(Y\|Z)\le\rho\alpha$ and $\mathrm{D}_\alpha(Z\|Y)\le\rho\alpha$.
\end{definition}

This is the insertion/deletion Feature-DP of \cite{fdp} (Def.~6) with R\'enyi divergence in place of the $(\varepsilon,\delta)$ inequalities, as zCDP relates to DP. The privacy unit is one record $z$, i.e., one text with its attribute values. The usual conversion from zCDP to $(\varepsilon,\delta)$-DP, e.g., $\varepsilon=\rho+2\sqrt{\rho\ln(1/\delta)}$~\cite[Prop.~1.3]{bun2016concentrated}, uses only these two divergence bounds. Applied to $(Y,Z)$ and $(Z,Y)$, it shows that $\rho$-feature-zCDP implies $(\varepsilon,\delta)$-Feature-DP (their $(\varepsilon,\delta)$-$\mathrm{DP}^{\Psi}_i$) for every $\delta\in(0,1)$.

\begin{theorem}
\label{thm:e2e}\label{thm:aim}
\sdgci{} (Algorithm~\ref{alg:sdg_ci}) with $\rho=\rho_{\mathrm{AIM}}+\rho_{\mathrm{PE}}$ is $\rho$-feature-zCDP relative to $\Psi_j$, and hence $(\varepsilon,\delta)$-Feature-DP for the conversion of $\rho$. Its Conditional AIM stage (Algorithm~\ref{alg:cond_aim}) is $\rho_{\mathrm{AIM}}$-feature-zCDP relative to $\Psi_j$, and, given the output $\widehat T$ of that stage, its PE stage is $\rho_{\mathrm{PE}}$-zCDP, i.e., standard DP on the whole text, which implies $\rho_{\mathrm{PE}}$-feature-zCDP.
\end{theorem}

\begin{proof}
\emph{Structure.} Write $P=\Psi_j(D_{\mathrm{in}})=\{\Psi_j(z'):z'\in D_{\mathrm{in}}\}$ for the multiset of public parts of the input corpus $D_{\mathrm{in}}$, with $n=|P|$. \sdgci{} accesses $D_{\mathrm{in}}$ only through $P$, which it copies exactly (Algorithm~\ref{alg:cond_aim}, line~\ref{line:copy}), and an adaptive sequence of noisy statistics $m_1,\ldots,m_M$. The $k$-th statistic $m_k(\cdot\mid P,m_{1:k-1})$ may be chosen based on $P$ and the earlier noisy outputs $m_{1:k-1}$. The noisy statistics are the one-way marginals of $\Apriv(c_j)$ (line~\ref{line:oneway}). In each AIM round, the marginal selected by the exponential mechanism and its noisy measurement (lines~\ref{line:select} and~\ref{line:measure}), and the $T_{\mathrm{PE}}+1$ noisy vote histograms of PE (Algorithm~\ref{alg:sdg_ci}, line~\ref{line:noise}), whose candidate pools are generated from $\widehat T$ and hence from $P$. PE does not use the public/private split, instead each record votes with its whole text $t$, so its histograms are protected like any other noisy statistic. All remaining steps, including the adaptive choice of $(\varepsilon_t,\sigma_t)$ in AIM, use only $P$, $m_{1:M}$, the fixed schema $\mathcal S$ and value sets $\mathcal V$ (A1, A4), and public in-context examples (A3, A5). Hence, for a fixed randomized function $G$,
\begin{align*}
\mathcal M(D_{\mathrm{in}})=G\big(P,\,&m_1(D_{\mathrm{in}}\mid P),\ldots,\\
&m_M(D_{\mathrm{in}}\mid P,m_{1:M-1})\big).
\end{align*}

\emph{Simulator.} Following the insertion/deletion definition of \cite{fdp}, the simulator receives $D$ and $\Psi_j(z)$, forms the public multiset $P=\Psi_j(D)\cup\{\Psi_j(z)\}=\Psi_j(D\cup\{z\})$ of size $n=|D|+1$, and outputs
\begin{align*}
\mathrm{sim}(D,\Psi_j(z))=G\big(P,\,&m_1(D\mid P),\ldots,\\
&m_M(D\mid P,m_{1:M-1})\big).
\end{align*}
That is, it runs the same algorithm with the same $P$ and the same record-count parameters, such as $\delta=1/n$ and the total count to which $\hat p_{t-1}$ is scaled (either $n$ or an estimate from earlier noisy outputs), but evaluates every noisy statistic on $D$ instead of $D\cup\{z\}$. The simulator never accesses $x[\Apriv(c_j)]$ or $t$.

\emph{Sensitivity.} Fix $P=\Psi_j(D\cup\{z\})$ and the earlier noisy outputs $m_{1:k-1}$. This fixes the marginal $M_r$ or the candidate pool of $h$, the scaled model $\hat p_{t-1}$, and the round parameters $(\varepsilon_t,\sigma_t)$, so the $k$-th statistic is the same query for $D\cup\{z\}$ and $D$. Let $\Delta M_r=M_r(D\cup\{z\})-M_r(D)$ for a measured AIM marginal $M_r$, and $\Delta h=h(D\cup\{z\})-h(D)$ for a PE vote histogram $h$. Since $z$ adds one to exactly one cell of $M_r$ and casts a single unit vote in $h$,
\begin{equation*}
\|\Delta M_r\|_2=\|\Delta M_r\|_1=1,\qquad \|\Delta h\|_2=1.
\end{equation*}
Marginals over public attributes alone are never measured (line~\ref{line:workload}), since they are determined by $P$. For the selection score $q_r(\cdot)=w_r\big(\|M_r(\cdot)-M_r(\hat p_{t-1})\|_1-\sqrt{2/\pi}\,\sigma_t n_r\big)$, the reverse triangle inequality gives
\begin{align*}
|q_r(D\cup\{z\})-q_r(D)|\le w_r\|\Delta M_r\|_1\\
=w_r\le\max\nolimits_r w_r .
\end{align*}

The score can move in either direction when $z$ is added, so the exponential mechanism samples with probability $\propto\exp\big(\varepsilon_t q_r/(2\max_r w_r)\big)$. It is then $\varepsilon_t$-DP with bounded range, and hence $(\varepsilon_t^2/8)$-zCDP~\cite{cesar2021bounding}, as in AIM~\cite{mckenna_aim}.

\emph{Per-step bounds.} Writing $m_k^{+}=m_k(D\cup\{z\}\mid P,m_{1:k-1})$ and $m_k^{-}=m_k(D\mid P,m_{1:k-1})$, each step satisfies
\begin{equation*}
\max\big\{\mathrm{D}_\alpha(m_k^{+}\|m_k^{-}),\,\mathrm{D}_\alpha(m_k^{-}\|m_k^{+})\big\}\le\alpha\rho_k,
\end{equation*}
with $\rho_k=1/(2\sigma_k^2)$ for a Gaussian measurement with sensitivity~1~\cite{bun2016concentrated} and $\rho_k=\varepsilon_t^2/8$ for a selection. Since $(\varepsilon_t,\sigma_t)$, and hence $\rho_k$, is a function of $m_{1:k-1}$, and the budget cap is enforced on the realized sum, adaptive composition applies.

\emph{Budget.} With $K=|\Apriv(c_j)|$ one-way marginals at noise $\sigma_0$, AIM rounds $t$ at $(\varepsilon_t,\sigma_t)$, and $T_{\mathrm{PE}}+1$ histograms at $\sigma_{\mathrm{PE}}=\sqrt{(T_{\mathrm{PE}}+1)/(2\rho_{\mathrm{PE}})}$,
\begin{align*}
\sum_{k=1}^{M}\rho_k
&=\underbrace{\frac{K}{2\sigma_0^2}}_{\le\,0.1\rho_{\mathrm{AIM}}}
+\underbrace{\sum_t\Big(\frac{\varepsilon_t^2}{8}+\frac{1}{2\sigma_t^2}\Big)}_{\le\,0.9\rho_{\mathrm{AIM}}}\\
&\quad+\underbrace{(T_{\mathrm{PE}}+1)\,\frac{1}{2\sigma_{\mathrm{PE}}^2}}_{=\,\rho_{\mathrm{PE}}}
\;\le\;\rho_{\mathrm{AIM}}+\rho_{\mathrm{PE}}=\rho .
\end{align*}
The first bound holds by the choice of $\sigma_0$ (line~\ref{line:oneway}), the second because line~\ref{line:final} shrinks the last round to the remainder, so the loop never exceeds $0.9\rho_{\mathrm{AIM}}$.

\emph{Composition and post-processing.} By the data-processing inequality for R\'enyi divergence and adaptive composition~\cite[Lemma~2.3]{bun2016concentrated},
\begin{align*}
\mathrm{D}_\alpha(Y\,\|\,Z)
&=\mathrm{D}_\alpha\Big(G\big(P,m^{+}_{1:M}\big)\,\Big\|\\
&\qquad\quad G\big(P,m^{-}_{1:M}\big)\Big)\\
&\le\mathrm{D}_\alpha\big(m^{+}_{1:M}\,\|\,m^{-}_{1:M}\big)\\
&\le\alpha\sum_{k=1}^{M}\rho_k\;\le\;\rho\alpha ,
\end{align*}
and the same holds for $\mathrm{D}_\alpha(Z\,\|\,Y)$.
This proves the claim for \sdgci{}. Restricting $G$ and the sum to Conditional AIM proves the claim for that stage. For the PE stage, the PE histograms are the only statistics PE computes from $D$, and comparing $D\cup\{z\}$ with $D$ directly, without a simulator, gives the standard zCDP bound, with total cost $\rho_{\mathrm{PE}}$ from the budget above. Standard zCDP implies feature-zCDP for any $\Psi_j$, since a simulator may ignore $\Psi_j(z)$~\cite[Prop.~2]{fdp}.
\end{proof}

Because $Z$ is produced without $t$ or $x[\Apriv(c_j)]$, the theorem bounds what the release reveals about the whole record beyond $\Psi_j(z)$, including content the schema does not capture. The guarantee is conditional on $\mathcal S$ and $\mathcal V$ (A1, A4). In our experiments both are derived from $D$ without DP, so which attributes and which specific values occur in the dataset as a whole is not protected. It does not limit inference from $\Psi_j(z)$ itself, which is released exactly, or membership inference (Limitations). For instance, the number of output rows reveals $|D|+1$, which the simulator is allowed to know. It does not apply to \anci{}, which has no formal guarantee.
 
 
\section{Hyperparameters}
\label{app:hyperparameters}
 
Table~\ref{tab:hyperparams} lists the configuration used for every \textsc{SDG-CI} and \textsc{ANON-CI} run unless stated otherwise. At $\varepsilon=\infty$ the generator receives $\rho_\text{AIM}=1000$ and the PE histogram is computed without adding noise.
 
\begin{table}[!ht]
\centering\small
\setlength{\tabcolsep}{4pt}
\begin{tabular}{@{}lp{0.5\columnwidth}@{}}
\toprule
Parameter & Value \\
\midrule
\multicolumn{2}{@{}l}{\textit{Budget}} \\
$\delta$ & $1/n$ ($n$ = number of private training records) \\
split & $\rho_\text{AIM}=0.1\rho$, $\rho_\text{PE}=0.9\rho$ \\
\midrule
\multicolumn{2}{@{}l}{\textit{Conditional AIM}} \\
workload $W$ & all cliques with $\le 3$ attributes and $\ge 1$ private attribute \\
max model size & 80\,MB (AIM default) \\
\midrule
\multicolumn{2}{@{}l}{\textit{PE (\textsc{SDG-CI})}} \\
LLM & Llama-3.1-8B-Instruct \\
embedder & sentence-t5-base \\
$T_\text{PE}$ & 2 \\
variations per candidate & 6 \\
\midrule
\multicolumn{2}{@{}l}{\textit{\Classify{}}} \\
LLM & Gemma-3-27B \\
\midrule
\multicolumn{2}{@{}l}{\textit{\textsc{ANON-CI}}} \\
LLM (all roles) & Llama-3.1-8B-Instruct \\
$p_\text{threshold}$ & 3 \\
$T_{\max}$ & 10 \\
\bottomrule
\end{tabular}
\caption{Hyperparameters of \textsc{SDG-CI} and \textsc{ANON-CI}.}
\label{tab:hyperparams}
\end{table}
 
\section{Audiences and Attribute Splits}
\label{app:audience_conditioned}
 
Table~\ref{tab:attributes} and Table~\ref{tab:audiences_datasets} specify the default attributes and their private and public splits used during evaluation. Table~\ref{tab:audiences_reddit} has the three Reddit audiences used in Figure~\ref{fig:spider_audiences}.
 
\begin{table*}[t]
\centering\small
\setlength{\tabcolsep}{4pt}
\begin{tabular}{@{}l p{0.78\textwidth}@{}}
\toprule
Dataset & Attributes \\
\midrule
Reddit &
age, city\_country, birth\_city\_country, income, relationship\_status, sex, education, occupation, income\_level, topic, word\_count. \\
\addlinespace
bioRxiv &
model\_organism, disease\_mention, sample\_size, category, primary\_research\_area, experimental\_approach, dominant\_data\_type, research\_focus\_scale, research\_goal, word\_count. \\
\addlinespace
RealEstate (SDG) &
location, home\_size, price\_home, preapproval\_status, conversation\_outcome, stage\_suggested, word\_count. \\
\addlinespace
RealEstate (Anon.) &
location, desired\_location, household, third\_parties\_mentioned, transaction\_stage, call\_purpose, buying\_amount, journey\_stage, next\_step. \\
\bottomrule
\end{tabular}
\caption{Attributes used for each dataset.}
\label{tab:attributes}
\end{table*}
 
\begin{table*}[t]
\centering
\footnotesize
\setlength{\tabcolsep}{4pt}
\renewcommand{\arraystretch}{1.15}
\begin{tabular}{@{}l p{0.25\textwidth} p{0.22\textwidth} p{0.22\textwidth} p{0.14\textwidth}@{}}
\toprule
Dataset & Information flow & Private (protected) & Public (released as is) & Utility label \\
\midrule
Reddit
 & \raggedright \textit{Sender:} Reddit users commenting on public forums\newline
   \textit{Recipient:} an external ML vendor\newline
   \textit{Purpose:} developing a content recommendation system on users' writing patterns
 & \raggedright age, city\_country, birth\_city\_country, income, relationship\_status
 & \raggedright sex, education, occupation, income\_level, topic, word\_count
 & \raggedright topic \tabularnewline
\midrule
bioRxiv
 & \raggedright \textit{Sender:} the research lab that conducted the experiments\newline
   \textit{Recipient:} the general public (open release)\newline
   \textit{Purpose:} disseminating experimental records so the findings can be read, reused and built upon
 & \raggedright model\_organism, disease\_mention, sample\_size
 & \raggedright category, primary\_research\_area, experimental\_approach, dominant\_data\_type, research\_focus\_scale, research\_goal, word\_count
 & \raggedright category \tabularnewline
\midrule
RealEstate
 & \raggedright \textit{Sender:} homebuyers and real estate agents\newline
   \textit{Recipient:} Internal Audit Team\newline
   \textit{Purpose:} assessing the fairness and compliance of the recorded conversations
 & \raggedright \textbf{SDG:} location, home\_size\newline\vspace{2pt}%
   \textbf{Anon.:} location, desired\_location, household, third\_parties\_mentioned, transaction\_stage
 & \raggedright \textbf{SDG:} price\_home, preapproval\_status, conversation\_outcome, stage\_suggested, word\_count\newline\vspace{2pt}%
   \textbf{Anon.:} call\_purpose, buying\_amount, journey\_stage, next\_step
 & \raggedright stage\_suggested \tabularnewline
\bottomrule
\end{tabular}
\caption{Default CI partition per dataset.}
\label{tab:audiences_datasets}
\end{table*}
 
\begin{table*}[t]
\centering
\footnotesize
\setlength{\tabcolsep}{4pt}
\renewcommand{\arraystretch}{1.15}
\begin{tabular}{@{}l p{0.31\textwidth} p{0.21\textwidth} p{0.21\textwidth}@{}}
\toprule
Split (priv/pub) & Downstream audience & Private (protected) & Public (released as is) \\
\midrule
5 / 6
 & \raggedright \textit{Recipient:} an ML vendor training a recommender model on the comments\newline
   \textit{Purpose:} training a content recommendation system on users' writing patterns
 & \raggedright age, city\_country, birth\_city\_country, income, relationship\_status
 & \raggedright sex, education, occupation, income\_level, topic, word\_count \tabularnewline
\midrule
3 / 8
 & \raggedright \textit{Recipient:} third parties re-sharing the comments\newline
   \textit{Purpose:} releasing the attributes users routinely self-disclose in public posts, withholding only demographic identifiers
 & \raggedright sex, age, birth\_city\_country
 & \raggedright city\_country, relationship\_status, income\_level, topic, education, occupation, income, word\_count \tabularnewline
\midrule
9 / 2
 & \raggedright \textit{Recipient:} an external advertising company building ad campaigns and user profiles\newline
   \textit{Purpose:} targeted advertising models
 & \raggedright age, sex, city\_country, birth\_city\_country, education, occupation, income, income\_level, relationship\_status
 & \raggedright topic, word\_count \tabularnewline
\bottomrule
\end{tabular}
\caption{CI partitions of the Reddit attributes for three downstream audiences.}
\label{tab:audiences_reddit}
\end{table*}
 
\section{Expert CI Classification Review}
\label{app:expert_ci_review}
We have asked a legal and domain expert to review the LLM-produced attribute split for our proprietary RealEstate dataset, constituting the hardest real-world example. For an external audience (Loan Officer), we obtained agreement on 7 out of 10 attributes (which is 9 anonymous attributes from Table~\ref{tab:attributes} + word\_count), with 3 attributes that were inferred from the dataset being classified as private by the LLM while the expert classified them as public (namely, transaction\_stage, journey\_stage and next\_step). For internal audiences (Product Quality and Internal Audit teams), we obtained agreement on 9 out of 10 attributes, with disagreement on word\_count, which was deemed public in nature and clarification on buying\_amount and desired\_location where the suggestion was to ensure that the coarsened version of the attributes would be used, e.g., metro or zip instead of the specific location and a price band instead of the specific amount.
 
\section{Attribute Inference Attacks}
\label{app:metrics}
We mount two different realistic attribute inference attacks against anonymization and SDG-based methods. Leak is computed as the mean over $a\in\Apriv(c_j)$ of the per-attribute score below.
 
\paragraph{SDG methods.}
We assume the attacker has access to the released synthetic data and knows the public attributes of a real target, and wants to infer the private attributes. The attacker then learns  $p(\Apriv\mid\Apub)$ by training a logistic regression classifier on the tabular representation of $\hat{D}$ and using it to infer $\Apriv$ in the held-out real test split. The score is the advantage over the majority class prediction,
\[
\mathrm{adv}_a=\mathrm{acc}_a-\mathrm{acc}_{\text{majority}}
\]
 
\paragraph{Anonymization methods.}
Here, the adversary sees the anonymized rewrite $\hat t_i$ of a real record and tries to infer its private attributes from a set of possible values \cite{staab2025language}. The prompt is ``Classify the $\langle a\rangle$ of this $\langle$domain$\rangle$. Reply with EXACTLY one of: $\langle$values$\rangle$'' \cite{rupta}. The score for $a$ is computed based on how often the adversary is able to infer the private attribute value.
 
 
\section{RealEstate Results}
\label{app:realestate}
 
\begin{figure*}[t]
\centering
\definecolor{sdgA}{rgb}{0.05,0.25,0.60}
\definecolor{sdgB}{rgb}{0.20,0.50,0.85}
\definecolor{sdgC}{rgb}{0.40,0.75,0.95}
\definecolor{anoA}{rgb}{0.70,0.10,0.10}
\definecolor{anoB}{rgb}{0.93,0.40,0.10}
\definecolor{anoC}{rgb}{0.96,0.70,0.20}
\tikzset{spider/.style={line width=0.9pt, fill opacity=0.05, mark size=1.3pt, mark options={solid, fill opacity=1}}}
\begin{subfigure}[t]{0.40\textwidth}
\centering
\resizebox{\linewidth}{!}{%
\begin{tikzpicture}[font=\footnotesize]
  \useasboundingbox (-2.75,-1.95) rectangle (2.75,2.15);
  \draw[gray!22] (90.00:0.425) -- (30.00:0.425) -- (-30.00:0.425) -- (-90.00:0.425) -- (-150.00:0.425) -- (-210.00:0.425) -- cycle;
  \node[gray, font=\tiny, inner sep=1pt] at (60.00:0.425) {0.25};
  \draw[gray!22] (90.00:0.850) -- (30.00:0.850) -- (-30.00:0.850) -- (-90.00:0.850) -- (-150.00:0.850) -- (-210.00:0.850) -- cycle;
  \node[gray, font=\tiny, inner sep=1pt] at (60.00:0.850) {0.5};
  \draw[gray!22] (90.00:1.275) -- (30.00:1.275) -- (-30.00:1.275) -- (-90.00:1.275) -- (-150.00:1.275) -- (-210.00:1.275) -- cycle;
  \node[gray, font=\tiny, inner sep=1pt] at (60.00:1.275) {0.75};
  \draw[gray!35] (90.00:1.700) -- (30.00:1.700) -- (-30.00:1.700) -- (-90.00:1.700) -- (-150.00:1.700) -- (-210.00:1.700) -- cycle;
  \node[gray, font=\tiny, inner sep=1pt] at (60.00:1.700) {1};
  \draw[gray!22] (0,0) -- (90.00:1.7);
  \node[anchor=270.00, inner sep=2pt] at (90.00:1.78) {MAUVE};
  \draw[gray!22] (0,0) -- (30.00:1.7);
  \node[anchor=210.00, inner sep=2pt] at (30.00:1.78) {1/(1+KL)};
  \draw[gray!22] (0,0) -- (-30.00:1.7);
  \node[anchor=150.00, inner sep=2pt] at (-30.00:1.78) {1$-$TV};
  \draw[gray!22] (0,0) -- (-90.00:1.7);
  \node[anchor=90.00, inner sep=2pt] at (-90.00:1.78) {1$-$JSD};
  \draw[gray!22] (0,0) -- (-150.00:1.7);
  \node[anchor=30.00, inner sep=2pt] at (-150.00:1.78) {Acc};
  \draw[gray!22] (0,0) -- (-210.00:1.7);
  \node[anchor=330.00, inner sep=2pt] at (-210.00:1.78) {Privacy};
  \draw[sdgA, solid, mark=*, spider, fill=sdgA] plot coordinates {(90.00:0.383) (30.00:0.476) (-30.00:0.709) (-90.00:1.423) (-150.00:0.556) (-210.00:1.574) (90.00:0.383)};
  \draw[sdgB, dashed, mark=square*, spider, fill=sdgB] plot coordinates {(90.00:0.218) (30.00:0.369) (-30.00:0.666) (-90.00:1.486) (-150.00:0.442) (-210.00:1.700) (90.00:0.218)};
  \draw[sdgC, densely dotted, mark=triangle*, spider, fill=sdgC] plot coordinates {(90.00:0.017) (30.00:0.143) (-30.00:0.126) (-90.00:1.035) (-150.00:0.532) (-210.00:1.689) (90.00:0.017)};
\end{tikzpicture}}
\caption{SDG}
\label{fig:spider-re-sdg}
\end{subfigure}\hspace{0.04\textwidth}
\begin{subfigure}[t]{0.40\textwidth}
\centering
\resizebox{\linewidth}{!}{%
\begin{tikzpicture}[font=\footnotesize]
  \useasboundingbox (-2.75,-1.95) rectangle (2.75,2.15);
  \draw[gray!22] (90.00:0.425) -- (0.00:0.425) -- (-90.00:0.425) -- (-180.00:0.425) -- cycle;
  \node[gray, font=\tiny, inner sep=1pt] at (45.00:0.425) {0.25};
  \draw[gray!22] (90.00:0.850) -- (0.00:0.850) -- (-90.00:0.850) -- (-180.00:0.850) -- cycle;
  \node[gray, font=\tiny, inner sep=1pt] at (45.00:0.850) {0.5};
  \draw[gray!22] (90.00:1.275) -- (0.00:1.275) -- (-90.00:1.275) -- (-180.00:1.275) -- cycle;
  \node[gray, font=\tiny, inner sep=1pt] at (45.00:1.275) {0.75};
  \draw[gray!35] (90.00:1.700) -- (0.00:1.700) -- (-90.00:1.700) -- (-180.00:1.700) -- cycle;
  \node[gray, font=\tiny, inner sep=1pt] at (45.00:1.700) {1};
  \draw[gray!22] (0,0) -- (90.00:1.7);
  \node[anchor=270.00, inner sep=2pt] at (90.00:1.78) {MAUVE};
  \draw[gray!22] (0,0) -- (0.00:1.7);
  \node[anchor=180.00, inner sep=2pt] at (0.00:1.78) {1/(1+KL)};
  \draw[gray!22] (0,0) -- (-90.00:1.7);
  \node[anchor=90.00, inner sep=2pt] at (-90.00:1.78) {1$-$TV};
  \draw[gray!22] (0,0) -- (-180.00:1.7);
  \node[anchor=0.00, inner sep=2pt] at (-180.00:1.78) {Privacy};
  \draw[anoA, solid, mark=*, spider, fill=anoA] plot coordinates {(90.00:1.630) (0.00:1.525) (-90.00:1.428) (-180.00:1.020) (90.00:1.630)};
  \draw[anoB, dashed, mark=square*, spider, fill=anoB] plot coordinates {(90.00:1.187) (0.00:1.019) (-90.00:1.243) (-180.00:1.028) (90.00:1.187)};
  \draw[anoC, densely dotted, mark=triangle*, spider, fill=anoC] plot coordinates {(90.00:1.032) (0.00:1.166) (-90.00:1.165) (-180.00:1.062) (90.00:1.032)};
\end{tikzpicture}}
\caption{Anonymization}
\label{fig:spider-re-anon}
\end{subfigure}
\par\medskip
{\footnotesize
\textbf{SDG} ($\varepsilon=1$):\quad \tikz[baseline=-0.6ex]\draw[sdgA, solid, line width=0.9pt, mark=*, mark size=1.3pt, mark options={solid}] plot coordinates {(0,0) (0.35,0) (0.7,0)};~SDG-CI\quad \tikz[baseline=-0.6ex]\draw[sdgB, dashed, line width=0.9pt, mark=square*, mark size=1.3pt, mark options={solid}] plot coordinates {(0,0) (0.35,0) (0.7,0)};~MAPLE\quad \tikz[baseline=-0.6ex]\draw[sdgC, densely dotted, line width=0.9pt, mark=triangle*, mark size=1.3pt, mark options={solid}] plot coordinates {(0,0) (0.35,0) (0.7,0)};~Aug-PE\\[2pt]
\textbf{Anonymization}:\quad \tikz[baseline=-0.6ex]\draw[anoA, solid, line width=0.9pt, mark=*, mark size=1.3pt, mark options={solid}] plot coordinates {(0,0) (0.35,0) (0.7,0)};~ANON-CI\quad \tikz[baseline=-0.6ex]\draw[anoB, dashed, line width=0.9pt, mark=square*, mark size=1.3pt, mark options={solid}] plot coordinates {(0,0) (0.35,0) (0.7,0)};~RUPTA\quad \tikz[baseline=-0.6ex]\draw[anoC, densely dotted, line width=0.9pt, mark=triangle*, mark size=1.3pt, mark options={solid}] plot coordinates {(0,0) (0.35,0) (0.7,0)};~1-2-3 Check
}
\caption{RealEstate results. All axes are higher-is-better. SDG and anonymization use different attribute splits (Appendix~\ref{app:realestate}).}
\label{fig:spider_realestate}
\end{figure*}
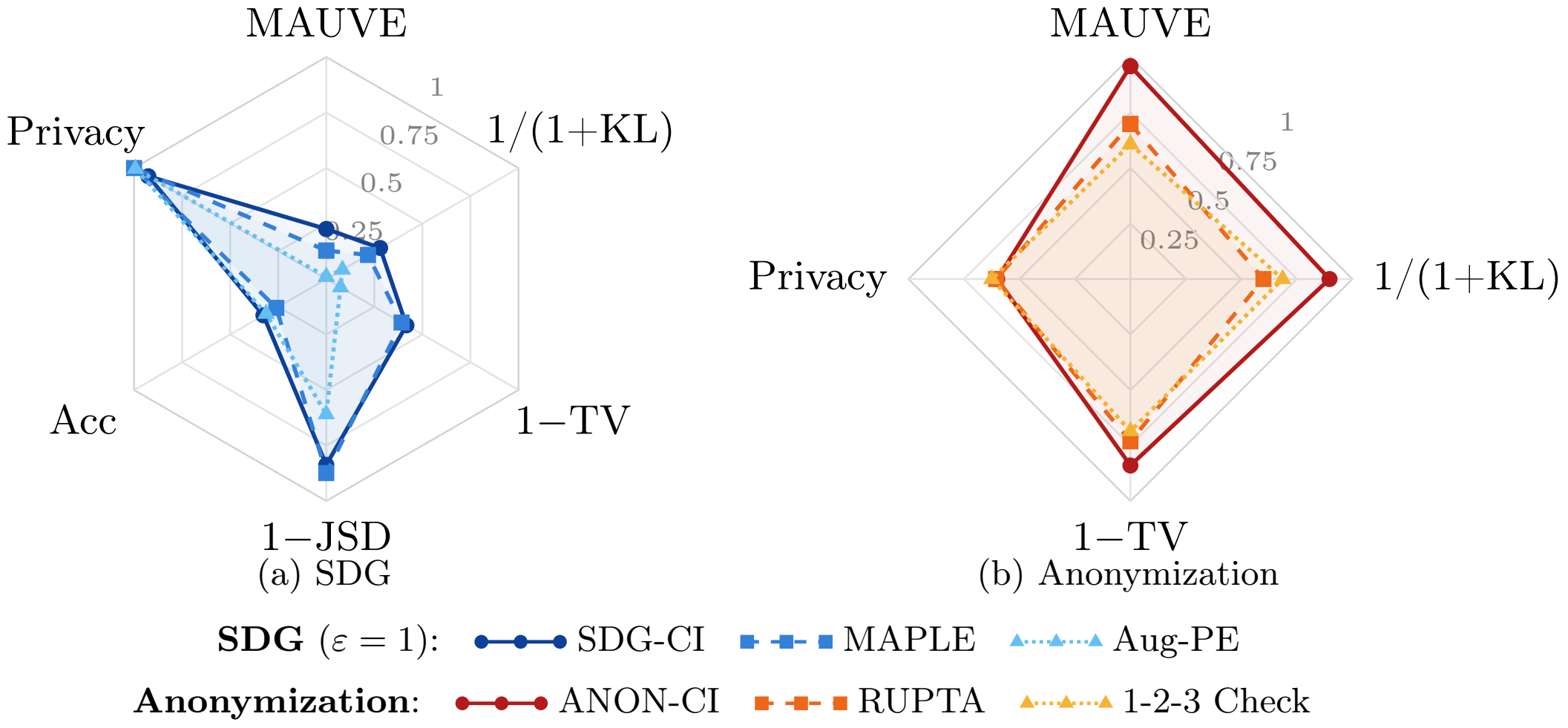

RealEstate is a proprietary conversational dataset in real estate domain with the user's home-buying stage as the utility label. Here, we investigate different attribute schemas for the ANON-CI and SDG-CI. The results in Figure~\ref{fig:spider_realestate} are computed using the SDG-CI and ANON-CI attribute schemas tailored for Internal Audit audience as described in Table~\ref{tab:audiences_datasets}.
 
 
 
\section{Exact Results}
\label{app:ablation}
Tables~\ref{tab:spider_audiences}, \ref{tab:spider_datasets}, and \ref{tab:spider_realestate} list the exact values behind Figures~\ref{fig:spider_audiences} and~\ref{fig:spider_realestate}.
 
\begin{table*}[t]
\centering
\small
\caption{Exact values for Figure~\ref{fig:spider_audiences}(a-c). SDG methods at $\varepsilon=1$. Best value per method family in bold.}
\label{tab:spider_audiences}
\begin{tabular}{llrrrrrrr}
\toprule
 & Method & MAUVE$\uparrow$ & KL$\downarrow$ & TV$\downarrow$ & JSD$\downarrow$ & Acc$\uparrow$ & Demographic Gap$\downarrow$ & Leak$\downarrow$ \\
\midrule
\multicolumn{9}{l}{\textit{ML Vendor}} \\
SDG & SDG-CI & \textbf{0.479} & \textbf{0.791} & \textbf{0.422} & \textbf{0.099} & \textbf{0.546} & 0.038 & 0.004 \\
 & MAPLE & 0.350 & 1.178 & 0.535 & 0.257 & 0.527 & \textbf{0.003} & \textbf{0.000} \\
 & Aug-PE & 0.027 & 10.557 & 0.814 & 0.188 & 0.522 & 0.084 & 0.001 \\
\cmidrule(l){2-9}
Anon. & ANON-CI & \textbf{1.000} & \textbf{0.002} & 0.019 & \textbf{0.011} & \textbf{0.734} & \textbf{0.013} & \textbf{0.111} \\
 & RUPTA & \textbf{1.000} & \textbf{0.002} & \textbf{0.013} & 0.016 & 0.725 & 0.053 & 0.114 \\
 & 1-2-3 Check & 0.982 & 0.065 & 0.110 & 0.048 & 0.671 & 0.029 & 0.159 \\
\midrule
\multicolumn{9}{l}{\textit{Re-post on another channel}} \\
SDG & SDG-CI & \textbf{0.546} & \textbf{0.569} & \textbf{0.407} & \textbf{0.054} & 0.507 & 0.015 & 0.018 \\
 & MAPLE & 0.350 & 1.178 & 0.535 & 0.257 & \textbf{0.527} & \textbf{0.003} & \textbf{0.000} \\
 & Aug-PE & 0.027 & 10.557 & 0.814 & 0.188 & 0.522 & 0.084 & 0.035 \\
\cmidrule(l){2-9}
Anon. & ANON-CI & \textbf{1.000} & \textbf{0.001} & 0.016 & \textbf{0.008} & \textbf{0.749} & \textbf{0.003} & \textbf{0.339} \\
 & RUPTA & \textbf{1.000} & 0.002 & \textbf{0.013} & 0.016 & 0.725 & 0.053 & 0.344 \\
 & 1-2-3 Check & 0.982 & 0.065 & 0.110 & 0.048 & 0.671 & 0.029 & 0.358 \\
\midrule
\multicolumn{9}{l}{\textit{Advertisement Company}} \\
SDG & SDG-CI & \textbf{0.400} & \textbf{0.704} & \textbf{0.495} & \textbf{0.162} & \textbf{0.604} & 0.062 & 0.003 \\
 & MAPLE & 0.350 & 1.178 & 0.535 & 0.257 & 0.527 & \textbf{0.003} & 0.001 \\
 & Aug-PE & 0.027 & 10.557 & 0.814 & 0.188 & 0.522 & 0.084 & \textbf{0.000} \\
\cmidrule(l){2-9}
Anon. & ANON-CI & \textbf{1.000} & 0.004 & 0.028 & \textbf{0.013} & \textbf{0.729} & 0.062 & \textbf{0.235} \\
 & RUPTA & \textbf{1.000} & \textbf{0.002} & \textbf{0.013} & 0.016 & 0.725 & 0.053 & 0.257 \\
 & 1-2-3 Check & 0.982 & 0.065 & 0.110 & 0.048 & 0.671 & \textbf{0.029} & 0.282 \\
\bottomrule
\end{tabular}
\end{table*}

\begin{table*}[t]
\centering
\small
\caption{Exact values for Figure~\ref{fig:spider_datasets}(a,d). The demographic gap is reported for Reddit only. SDG methods at $\varepsilon=1$. Best value per method family in bold.}
\label{tab:spider_datasets}
\begin{tabular}{llrrrrrrr}
\toprule
 & Method & MAUVE$\uparrow$ & KL$\downarrow$ & TV$\downarrow$ & JSD$\downarrow$ & Acc$\uparrow$ & Demographic Gap$\downarrow$ & Leak$\downarrow$ \\
\midrule
\multicolumn{9}{l}{\textit{Reddit}} \\
SDG & SDG-CI & \textbf{0.479} & \textbf{0.791} & \textbf{0.422} & \textbf{0.099} & \textbf{0.546} & 0.038 & 0.004 \\
 & MAPLE & 0.350 & 1.178 & 0.535 & 0.257 & 0.527 & \textbf{0.003} & \textbf{0.000} \\
 & Aug-PE & 0.027 & 10.557 & 0.814 & 0.188 & 0.522 & 0.084 & 0.001 \\
\cmidrule(l){2-9}
Anon. & ANON-CI & \textbf{1.000} & \textbf{0.002} & 0.019 & \textbf{0.011} & \textbf{0.734} & \textbf{0.013} & \textbf{0.111} \\
 & RUPTA & \textbf{1.000} & \textbf{0.002} & \textbf{0.013} & 0.016 & 0.725 & 0.053 & 0.114 \\
 & 1-2-3 Check & 0.982 & 0.065 & 0.110 & 0.048 & 0.671 & 0.029 & 0.159 \\
\midrule
\multicolumn{9}{l}{\textit{bioRxiv}} \\
SDG & SDG-CI & \textbf{0.420} & \textbf{1.750} & \textbf{0.463} & \textbf{0.124} & 0.746 & -- & \textbf{0.000} \\
 & MAPLE & 0.097 & 3.217 & 0.689 & 0.444 & 0.556 & -- & \textbf{0.000} \\
 & Aug-PE & 0.028 & 6.417 & 0.791 & 0.405 & \textbf{0.810} & -- & 0.023 \\
\cmidrule(l){2-9}
Anon. & ANON-CI & 0.999 & 0.019 & 0.077 & 0.047 & 0.423 & -- & \textbf{0.360} \\
 & RUPTA & \textbf{1.000} & \textbf{0.011} & \textbf{0.037} & \textbf{0.030} & \textbf{0.866} & -- & 0.380 \\
 & 1-2-3 Check & 0.998 & 0.024 & 0.077 & 0.053 & 0.387 & -- & 0.439 \\
\bottomrule
\end{tabular}
\end{table*}

\begin{table*}[t]
\centering
\small
\caption{Exact values for Figure~\ref{fig:spider_realestate}. SDG methods at $\varepsilon=1$. Best value per method family in bold.}
\label{tab:spider_realestate}
\begin{tabular}{llrrrrrr}
\toprule
 & Method & MAUVE$\uparrow$ & KL$\downarrow$ & TV$\downarrow$ & JSD$\downarrow$ & Acc$\uparrow$ & Leak$\downarrow$ \\
\midrule
\multicolumn{8}{l}{\textit{RealEstate}} \\
SDG & SDG-CI & \textbf{0.225} & \textbf{2.570} & \textbf{0.583} & 0.163 & \textbf{0.327} & 0.074 \\
 & MAPLE & 0.128 & 3.608 & 0.608 & \textbf{0.126} & 0.260 & \textbf{0.000} \\
 & Aug-PE & 0.010 & 10.928 & 0.926 & 0.391 & 0.313 & 0.007 \\
\cmidrule(l){2-8}
Anon. & ANON-CI & \textbf{0.959} & \textbf{0.115} & \textbf{0.160} & -- & -- & 0.400 \\
 & RUPTA & 0.698 & 0.668 & 0.269 & -- & -- & 0.395 \\
 & 1-2-3 Check & 0.607 & 0.458 & 0.315 & -- & -- & \textbf{0.375} \\
\bottomrule
\end{tabular}
\end{table*}

\section{SDG Results across $\varepsilon$}
\label{app:epsilon_results}
Table~\ref{tab:eps_sweep} reports the performance of SDG methods across different privacy budgets on Reddit and bioRxiv datasets.
 
\begin{table*}[t]
\centering
\small
\setlength{\tabcolsep}{3.5pt}
\caption{SDG methods across privacy budgets $\varepsilon$. The demographic gap is reported for Reddit only. Best value per $\varepsilon$ in bold.}
\label{tab:eps_sweep}
\resizebox{\textwidth}{!}{%
\begin{tabular}{ll|rrr|rrr|rrr|rrr}
\toprule
 & \multicolumn{3}{c}{MAUVE$\uparrow$} & \multicolumn{3}{c}{Acc$\uparrow$} & \multicolumn{3}{c}{Demographic Gap$\downarrow$} & \multicolumn{3}{c}{Leak$\downarrow$} \\
Dataset & Method & 0.5 & 1 & 10 & 0.5 & 1 & 10 & 0.5 & 1 & 10 & 0.5 & 1 & 10 \\
\midrule
Reddit & SDG-CI & \textbf{0.520} & \textbf{0.479} & 0.464 & 0.502 & \textbf{0.546} & \textbf{0.638} & \textbf{0.014} & 0.038 & \textbf{0.022} & 0.004 & 0.004 & \textbf{0.000} \\
 & MAPLE & 0.302 & 0.350 & \textbf{0.711} & 0.478 & 0.527 & 0.551 & 0.051 & \textbf{0.003} & 0.048 & \textbf{0.000} & \textbf{0.000} & \textbf{0.000} \\
 & Aug-PE & 0.016 & 0.027 & 0.027 & \textbf{0.536} & 0.522 & 0.551 & 0.114 & 0.084 & 0.087 & \textbf{0.000} & 0.001 & 0.020 \\
\midrule
bioRxiv & SDG-CI & \textbf{0.194} & \textbf{0.420} & \textbf{0.668} & 0.662 & 0.746 & \textbf{0.796} & -- & -- & -- & \textbf{0.000} & \textbf{0.000} & 0.031 \\
 & MAPLE & 0.090 & 0.097 & 0.517 & 0.514 & 0.556 & 0.577 & -- & -- & -- & \textbf{0.000} & \textbf{0.000} & \textbf{0.000} \\
 & Aug-PE & 0.031 & 0.028 & 0.080 & \textbf{0.789} & \textbf{0.810} & 0.669 & -- & -- & -- & 0.019 & 0.023 & \textbf{0.000} \\
\bottomrule
\end{tabular}}
\end{table*}

\end{document}